\documentclass[prd,preprintnumbers,amsmath,amssymb,superscriptaddress]{revtex4}  
\usepackage{graphicx}
\usepackage{epsfig}
\usepackage{rotating}
\usepackage{dcolumn}
\usepackage{bm}
\newcommand{\lsim}{\raisebox{-0.13cm}{~\shortstack{$<$ \\[-0.07cm] $\sim$}}~} 
 
\newcommand{\beq}{\begin{eqnarray}} 
\newcommand{\eeq}{\end{eqnarray}}

\begin{document}

\preprint{LPT--ORSAY--13--19}

\title{{\Large\bf The couplings of the Higgs boson and its CP properties from fits}
\\Ê{\Large\bf of the signal strengths and their ratios at the 7+8 TeV LHC}}

\author{{\sc Abdelhak Djouadi} and {\sc Gr\'egory Moreau} \\
{\it Laboratoire de Physique Th\'eorique, B\^at. 210, CNRS,
Universit\'e Paris-sud 11 \\  F-91405 Orsay Cedex, France}}

\begin{abstract} 
Using the full set of the LHC Higgs data from the runs at 7 and 8 TeV center of
mass energies that have been  released by the ATLAS and CMS collaborations, we
determine the couplings of the Higgs particle to fermions and gauge bosons as
well as its  parity or CP composition. We consider  ratios of production cross
sections times decay branching fractions in which the theoretical (and some
experimental)  uncertainties as well as as some ambiguities from new physics
cancel out. A fit of both the signal strengths in the various search channels
that have been conducted, $H \to ZZ, WW, \gamma \gamma, \tau\tau$ and $b\bar b$,
and their  ratios shows that  the observed $\sim 126$ GeV particle has couplings to
fermions and gauge bosons that  are Standard Model--like already at the  68\%
confidence level (CL). From the signal strengths in which the theoretical 
uncertainty is taken to be a bias,  the particle is shown to  be at most 68\% CP--odd at the
99\%CL and the possibility that it is a pure pseudoscalar state is
excluded at the $4\sigma$ level when including both the  experimental and
theoretical uncertainties. The signal strengths also measure the invisible
Higgs decay width which, with the same type of uncertainty analysis, is shown to be 
$\Gamma_H^{\rm inv}/  \Gamma_H^{\rm SM} \leq 0.52$ at the 68\%CL.
\end{abstract}

\maketitle

\large 

\subsection*{{\Large 1. Introduction}}

The ATLAS and CMS collaborations have released their analyses of the mass and
the production times decay rates of the 126 GeV Higgs--like particle  using the
full set of data collected so far,  $\approx 5$ fb$^{-1}$ at $\sqrt s=7$ TeV and
$\approx 20$ fb$^{-1}$ at 
$\sqrt s=8$~TeV~\cite{CONF-2013-014,CONF-2013-034,PAS-HIG-12-045,PAS-HIG-13-001}. This closes a very
successful first  run at the  LHC, which culminated with the historical
discovery of the state in July 2012~\cite{ATLAS-disc,CMS-disc}. To be convinced
that the observed particle is indeed the Higgs boson that is responsible of the
spontaneous breaking of the electroweak symmetry~\cite{Higgs,Review}, one needs
to prove that the particle: $i)$ has spin--zero, $ii)$  is a CP--even  state,
$iii)$ couples to fermions and gauge bosons proportionally to their masses and,
ultimately,  $iv)$ has a self--coupling that is also proportional to its mass.
While there is little doubt on the spin--zero nature of the observed
state~{$^1$}\footnotetext[1]{The observation of the $H\to \gamma\gamma$ decay rules out the
spin--1 case~\cite{Landau-Yang} and the graviton--like spin--2 possibility  is
extremely unlikely and, from the particle rates, is ruled out in large classes
of models~\cite{Ellis}.}, and the probing of the self--coupling has to await 
for   a high--luminosity LHC~\cite{triple} or a  future lepton collider, a first
determination of the couplings~\cite{Fits,EFIT,Fit-last} and the 
CP--properties~\cite{CPH-d,CPH-p} can be  performed with the current results.   The ATLAS and
CMS collaborations themselves have given a first ``portrait" of the observed
particle which indicates that indeed it has the properties of a Higgs boson and
even more,  the properties of the unique Higgs boson that is predicted in the
Standard Model (SM)~\cite{CONF-2013-014,CONF-2013-034,PAS-HIG-12-045,PAS-HIG-13-001,CMS1212}. 

However, in the case of the particle coupling determination, the ATLAS and CMS 
analyses suffer from two serious drawbacks~\cite{ratios,Dieter}. The first one
is that the signal strength modifiers $\mu_{XX}$, that are  identified with the 
Higgs cross section times decay branching ratio normalized to the  SM
expectation  in a given $H\!\to\! XX$ search channel,  are affected by large
theoretical uncertainties that are now becoming a dominant source of error.
Indeed, the  combined theoretical uncertainty in the rate of the by far dominant
Higgs production  process at the LHC, gluon fusion  $gg\to H$, is estimated to
be of order $\pm 15$--20\%~\cite{LHCXS,BD} even before it is broken into  jet
categories which significantly increases the 
uncertainty~\cite{Scale-H0j,Scale-H2j}. The uncertainty is similar in the vector boson
fusion channel when the  large contamination from the $gg \to Hjj$ process is
taken into account~\cite{Scale-H2j}. Another drawback of the 
analyses~\cite{ratios,Dieter} is that they involve strong theoretical assumptions on the
total Higgs width since some contributing decay channels  not accessible at the
LHC are assumed  to be SM--like and possible invisible Higgs decays in scenarios
beyond the SM are supposed not to occur. 

In this letter, we consider ratios of Higgs production cross sections times
decay bran\-ching fractions in which these two sources of uncertainties and,
eventually, also some systematical and parametrical uncertainties such as the
error on the luminosity measurement and the one on the Higgs branching 
ratios~{$^2$}\footnotetext[2]{There are parametric uncertainties that affect the hadronic 
Higgs decay widths. For a $\approx 126$ GeV SM Higgs boson,  this translates
into an uncertainty on the total Higgs width which is found  to be of order
$\approx 4\%$ in Ref.~\cite{BRpaper} and slightly higher  in Ref.~\cite{BD}.}~\cite{BD,BRpaper},  
should be absent~\cite{ratios,Dieter}. Using the Higgs
signal strengths in which the theoretical uncertainty is taken to be a bias  and
not a nuisance (as is done generally),  as well as the ratios
$\mu_{\gamma\gamma}/\mu_{ZZ}$ and  $\mu_{\tau \tau}/\mu_{WW}$ 
which are free from the ambiguities above, we perform a fit of the
latest ATLAS and CMS data and conclude that, already at the 68\% confidence
level (CL),  there is no deviation of the Higgs couplings to fermions and gauge
bosons from the SM expectation.

On an other front, the attempts made so far for the determination of the CP
nature of the particle  mainly exploit the kinematical features of the $H\to VV$
decays with $V=W,Z$~\cite{CPH-d} or the production in the vector boson fusion
$V^*V^* \to H$ and Higgs--strahlung $V^* \to VH$ processes~\cite{CPH-p}. Since a
CP--odd particle has not have tree level couplings to $VV$ states~{$^3$}\footnotetext[3]{The
effective $VV$ coupling of a pseudoscalar $A$ boson,  $\propto  M_V^2  V^{\mu
\nu} \widetilde{V}_{\mu \nu}$ with $\widetilde{V}_{\mu \nu}= \epsilon^{\mu \nu
\rho \sigma} V_{\rho \sigma}$,  should be generated through tiny  loop
corrections. To make this coupling as large as the SM tree--level  $HVV$
coupling, one needs a very low new physics scale that would spoil the precision
electroweak data.}, all these processes project out only the CP--even component
of the $HVV$ coupling~\cite{CP-tt} and the considered distributions can  be thus
only those of a $0^{++}$ state. 

We will show that a much better way to measure the CP composition of the
observed Higgs state is to consider the measured signal strength in the $H\to
VV$ decays; see also Ref.~\cite{Schwaller}. 
Using $\mu_{ZZ}$, we demonstrate  that, if the magnitude of the Higgs
couplings to fermions is as in the SM, the pure  CP--odd possibility is excluded
at the $4 \sigma$ level, irrespective  of the (mixed CP) Higgs couplings
to light fermions. 

Finally, assuming that the Higgs couplings to fermions and gauge bosons are
SM--like and using mainly the signal strength $\mu_{ZZ}$, one obtains a
limit on the rate for invisible Higgs decays, 
$\Gamma_H^{\rm inv}/  \Gamma_H^{\rm SM} \leq 0.52$ 
at the 68\%CL (see also Ref.~\cite{Fit-last,Belanger} for an 
indirect limit), that is stronger than the one obtained from  direct invisible
Higgs searches~\cite{portal,InvATLAS}.

\subsection*{{\Large 2. The signal strengths and their ratios}}

Let us first summarize the LHC Higgs data collected in the 2011 and 2012 runs
for the various  SM Higgs decay channels that have been searched for by the
ATLAS and CMS collaborations: $H \to ZZ^* \to 4\ell^\pm, \ H\to WW^* \to 2\ell
2\nu, \ H \to \gamma\gamma, \ H \to \tau^+ \tau^-$  and $H\to b \bar b$; we will
ignore the additional search channels $H \to \mu^+ \mu^-$ and $H \to Z\gamma$
(see e.g. Ref.~\cite{Vega})
for which the sensitivity is still too low. In most cases, the various Higgs
production channels have been used:  the by far dominant gluon--gluon fusion
mechanism $gg\to H$  (ggF) that has the large production rates but also the
subleading channels, vector boson fusion (VBF) $qq \to Hqq$ and Higgs--strahlung
(HV) $q\bar q \to HV$ with $V=W,Z$; the top quark associated $p\bar p\to t\bar t
H$ mechanism (ttH) has too low a cross section to be relevant.  At least the ggF
and VBF channels have been considered in the $H\to ZZ, WW, \gamma \gamma$ and
$H\to \tau^+\tau^-$ channels, while in the case $H\to \tau^+\tau^-$ and $H\to
WW^*\to 2\ell 2\nu$ decays, also the HV production mode in which  the $H \to
b\bar b$ decay has been searched for, has been  considered.

ATLAS and CMS have provided the signal strengths for the various final states
with a luminosity of, respectively,  $\approx 5$ fb$^{-1}$ for  the 2011 run at
$\sqrt s=7$ TeV and $\approx 20$ fb$^{-1}$ the 2012 run at $\sqrt s=8$ TeV.  We
will identify these $\mu$ values with the Higgs cross section times  decay
branching fractions normalized to the SM  expectation and, for the $H\to XX$
decay, one would have indeed in the narrow width approximation, 
\beq
\mu_{XX}\vert_{\rm th} \simeq \frac {\sigma( pp \to H \to XX)}{ \sigma( pp \to H
\to XX)|_{\rm SM}} \simeq   \frac {\sigma( pp \to H)\times {\rm BR} (H \to XX)}{
\sigma( pp \to H)|_{\rm SM} \times {\rm BR} (H \to XX)|_{\rm SM} } .
\label{mudef}  
\eeq  
From the experimental point of view (and in our fits), this
would correspond  to  
\beq 
\mu_{XX}\vert_{\rm exp} \simeq  \frac {N^{\rm evts}_{XX}}{ \epsilon \times
\sigma( pp \to H)|_{\rm SM} \times {\rm BR} (H \to XX)|_{\rm SM} \times {\cal
L}},  \label{muEXP}  
\eeq  
where $N^{\rm evts}_{XX}$ stands for the measured number of events in the  $H\to
XX$ search channel, $\epsilon$ denotes the selection efficiency and  ${\cal L}$
is the  luminosity.

In this paper, we consider the decay ratios $D_{XX}$  discussed in
Ref.~\cite{ratios} and defined as 
\beq
D_{XX}^{\rm p} = \frac {\sigma^{\rm p} ( pp \to H \to XX)}{ \sigma^{\rm p}
( pp \to H \to VV)} =  \frac {\sigma^{\rm p} ( pp \to H)\times {\rm BR} 
(H \to XX)}{ \sigma^{\rm p}( pp \to H) \times {\rm BR} (H \to VV)} =
\frac{\Gamma( H \to XX)}{ \Gamma ( H \to VV)} \label{ratio} 
\eeq 
for a specific production process ${\rm p= ggF, VBF, VH}$ or all  (for inclusive
production) and  for a given decay channel $H\to XX$ when the reference channel
$H\to VV$ with $V=W$ or/and $Z$ is used. In these ratios,  the cross sections\
$\sigma(pp \to H)$ and hence, their significant theoretical  uncertainties will
cancel out as discussed previously, leaving out only the ratio of decay
branching fractions  and hence of partial decay widths. Thus, the total decay
width which includes contributions from channels not under control such as
possible invisible Higgs decays, do  not appear in the decay ratios $D_{XX}^{\rm
p}$.  Some common experimental systematical uncertainties such as the one from
the luminosity measurement as well as the uncertainties in the Higgs decay 
branching ratios also cancel out. We are thus, in principle,  left with mostly
the statistical uncertainty and some systematical errors~{$^4$}\footnotetext[4]{The
theoretical and common systematical uncertainties will completely cancel out
only when the  \underline{same}  \  
selection cuts are applied for the different
final state topologies in a given production process (the selection efficiencies
should be $\epsilon_X^{\rm p}= k\, \epsilon_V^{\rm  p}$, $k$ being a
constant). This is obviously not the case in all the channels that we are 
considering here. We will assume, nevertheless, that this will be the case and
we consider that the remaining uncertainty in the ratio is, to a good
approximation, only of statistical nature. We hope that in the future,  with the
much larger data sample that is expected, the ATLAS and CMS  collaborations will
analyze the various search channels under the  same experimental conditions.}.

The ratios $D_{XX}$ involve, up to  kinematical factors, only the ratios  $\vert
c_X \vert^2/\vert c_V\vert^2$ of the reduced  couplings of the Higgs  boson to
the particles  $X$ and $V$ compared to the SM expectation, $c_X \equiv
g_{HXX}/g_{HXX}^{\rm SM}$. $d_{XX}$. For the SM Higgs boson with  a mass of
$M_H=125$ GeV, the kinematical factors  can be straightforwardly    obtained
using the  program {\tt HDECAY}~\cite{HDECAY} for the evaluation of the Higgs
branching ratios when the SM inputs recommended by the LHC Higgs working 
group~\cite{LHCXS} are adopted; they are given in  Ref.~\cite{ratios} for the various
normalisations (and, for simplicity, are then set to unity in that paper).

In practice, to take into account the fact that there are four different Higgs
production channels  with different topologies, the cross section part is more
involved and the ratio  $D_{XX}$ can be more precisely written as
\beq
D_{XX} \propto \frac {\epsilon^{gg}_X \sigma( gg \! \to\! H)\! + \!
\epsilon^{\rm VBF}_X \sigma(Hqq) \! + \! 
\epsilon^{HV}_X \sigma(HV) +  \epsilon^{t\bar tH}_X \sigma(t\bar tH)}
{\epsilon^{gg}_V \sigma( gg \! \to \! H)+ \epsilon^{\rm VBF}_V
\sigma(Hqq) +  \epsilon^{HV}_V \sigma(HV)
+  \epsilon^{t\bar tH}_V \sigma(t\bar tH)} 
\times \ 
\frac{\frac {\Gamma( H \to XX)}{\Gamma( H \to XX)|_{\rm SM}}}{\frac { \Gamma ( 
H \to VV)}{\Gamma( H \to VV)|_{\rm SM}}}
\label{Dapprox}
\eeq
where the $\epsilon_X^p$, provided by the ATLAS and CMS collaborations, denote
the experimental efficiencies to select the Higgs events in the $gg$, VBF, $HV$,
$t\bar tH$ production and $H\to XX$ decay channels (exclusive cut categories are
also considered). $D_{XX}$ is only proportional to the above expression due to
the presence of another identical ratio of cross sections but within the  
SM~{$^5$}\footnotetext[5]{Without derivative Higgs couplings, the  kinematics and selection
efficiencies are as in the SM.}. Nevertheless for almost the same
selection  efficiencies, and even for  $\epsilon_X^p = k \epsilon_V^p$ with $k$
a constant, the  expression eq.~(\ref{Dapprox}) simplifies to that in
eq.~(\ref{ratio1}).  

In fact, the decay ratios $D_{XX}$  can be simply written in terms of the 
signal strengths
\beq
D_{XX} \hat =  \frac {\mu_{XX}}{\mu_{VV}} 
\simeq  \frac {\frac {\sigma( pp \to H)\times {\rm BR} (H \to XX)}{ \sigma( pp \to H)|_{\rm
SM} \times {\rm BR} (H \to XX)|_{\rm SM} }}{\frac {\sigma( pp \to H)\times {\rm
BR} (H \to VV)}{ \sigma( pp \to H)|_{\rm
SM} \times {\rm BR} (H \to VV)|_{\rm SM} }} 
= \frac {\frac {{\rm BR} (H \to XX)}{{\rm BR} (H \to XX)|_{\rm SM} }}{\frac
{{\rm BR} (H \to VV)}{{\rm BR} (H \to VV)|_{\rm SM} }}  
= \frac{\frac {\Gamma( H \to XX)}{\Gamma( H \to XX)|_{\rm SM}}}{\frac { \Gamma (
H \to VV)}{\Gamma( H \to VV)|_{\rm SM}}} = 
\frac{ \vert c_X\vert ^2}{\vert c_V \vert^2} \label{ratio1} 
\eeq
where have used as normalisations in the ratios the channels~{$^6$}\footnotetext[6]{In fact, 
one can assume custodial symmetry and  use the combined $H\to WW$ and $H\to ZZ$
channels as a reference to increase the statistical accuracy of the 
normalization factor. The ratio $D_{ZZ}=\mu_{ZZ}/\mu_{WW}$ has been measured 
for instance  by the ATLAS collaboration to be $D_{ZZ} \simeq \vert c_Z \vert^2/\vert
c_W \vert^2 =1.6^{+0.8}_{-0.5}$~\cite{CONF-2013-034} (with the error expected
to be only  statistical), hence, supporting this approach.} $H\to VV$ with
$V\!=\!Z$ or $W$.

Nevertheless, performing the ratios of signal strengths leads to a loss of 
information and, in some case, we will need at least one signal strength to set
the normalisation. Rather than using the global $\mu_{\rm tot}$ value obtained
by combining all search channels, we will consider the cleaner  $H\to ZZ$
channel alone as  it is  fully inclusive and thus does not involve the
additional large scale uncertainties that occur when breaking the $gg\! \to\! H$
cross section into jet categories~{$^7$}\footnotetext[7]{In addition, contrary to the global
signal strength $\mu_{\rm tot}$, it does not involve the channel $\Gamma(H\to
\gamma\gamma)$ which, at least in the ATLAS case, deviates from the SM
prediction and  might indicate the presence of new physics in the 
$H\gamma\gamma$ loop which is not reflected in the other couplings.}. The
combination of the ATLAS and CMS data in the $ZZ$ channel gives
\begin{eqnarray}   
\mu_{\rm ZZ} =1.10 \pm 0.22^{\rm \; exp} \pm 0.2^{\rm \; th} \, 
\label{muZZ}  
\end{eqnarray}
where the first uncertainty is experimental and the second one theoretical and
that we assume to be, conservatively,  $\Delta^{\rm th}\!=\!\pm
0.2$. It has been advocated in Ref.~\cite{ratios,BD} that, since the main effect
of the  theoretical uncertainty (which has no statistical ground) is to modify
the normalisation of the SM cross section, it should be considered as a bias 
(rather than  a nuisance as in the case of the experimental error) and, hence,
one needs to perform two separate fits: one with  $\mu_{ZZ}\! +\! \Delta^{\rm
th}$ and another with $\mu_{ZZ}\! - \!\Delta^{\rm th}$.

\subsection*{{\Large 3. A combined fit of the Higgs couplings}}

In order to study the Higgs at the LHC we define the (now usual)  effective
Lagrangian,     
\begin{eqnarray} 
{\cal L}_h  & = &  \ c_W \ g_{HWW} \ H \ W_{\mu}^+ W^{- \mu} + \ c_Z \ g_{HZZ} \ H \ Z_{\mu}^0 Z^{0 \mu}
\label{Eq:LagEff}\\ &- &   
  c_t \; y_t\;  H  \bar t_L  t_R  -  c_c \; y_c \; H  \bar c_L  c_R  - 
  c_b \; y_b\;   H  \bar b_L b_R  - c_\tau \; y_\tau \; H  \bar \tau_L \tau_R 
\  + \ {\rm h.c.} \nonumber 
\end{eqnarray}
where $y_{t,c,b,\tau}=m_{t,c,b,\tau}/v$ are the SM Yukawa coupling constants in
the mass eigenbasis ($L/R$ indicates the fermion chirality and we consider  only
the heavy fermions that have substantial couplings to the Higgs boson),  
$g_{HWW} = 2M^2_W/v$ and $g_{HZZ} = M^2_Z/v$ are the electroweak gauge boson
couplings and $v$ is the Higgs vacuum expectation value.  The $c$ parameters are
all defined such that the limit $c\to 1$ corresponds to the SM case. For the
present task, we assume no or negligible new contributions to the Higgs
couplings to photons or gluons, e.g. as induced  by  new particles. 

We will present the results for the fits of the Higgs signal strengths, $\mu_i$
($i$ labels each channel and cut category investigated), in the plane $c_f$
versus $c_V$. We have chosen universal coupling corrections,
$c_f=c_t=c_c=c_b=c_\tau$ and $c_V=c_W=c_Z$,  for an illustrative purpose. All
the Higgs production/decay channels are considered here and the data used are
the latest ones and are borrowed from Ref.~\cite{muTevatron} for the Tevatron, from
Refs.~\cite{CONF-2013-012,CONF-2013-013,CONF-2013-030,CONF-2012-170,CONF-2012-160} 
and the combined results of  Refs.~\cite{CONF-2013-014,CONF-2013-034} for the ATLAS collaboration as well as   
Refs.~\cite{PAS-HIG-12-053,PAS-HIG-13-001,PAS-HIG-13-002,PAS-HIG-13-003,PAS-HIG-12-020,PAS-HIG-12-045,PAS-HIG-13-004} 
and the combined analysis of Ref.~\cite{PAS-HIG-12-045} for the CMS collaboration
and finally the LHC results presented at the Moriond conference this month~\cite{Ochando,InvMoriond}.

We will closely follow the procedure of Ref.~\cite{EFIT}  (to which we refer for
the relevant details) for the fit of these data, with one  major difference
though, the treatment of the theoretical uncertainty.   The errors used in the
present fit are those given by the collaborations as quoted above and, thus,
contain   the experimental and theoretical uncertainties added in quadrature,
$\delta\mu_i = \sqrt{\delta\mu_i\vert_{\rm exp}^2+\delta\mu_i\vert_{\rm th}^2}$,
which completely dilutes the effect of  $\delta\mu_i\vert_{\rm th}$. In the
present analysis,  we treat the theoretical uncertainty as a bias (and not as if
it were associated with a statistical distribution) and perform the fit for the
two extremal values of the  signal strength~: $\mu_{i} \vert_{ \rm exp} [ 1 \pm
\delta \mu_i/\mu_i \vert_{\rm th} ]$ with  the theoretical uncertainty
$\delta\mu_i/ \mu_i \vert_{\rm th}$  conservatively assumed to be $20\%$ for
both the gluon and vector boson fusion mechanisms.

\begin{figure}[t]
\begin{center}
\begin{tabular}{c}
\includegraphics[width=0.45\textwidth,height=8cm]{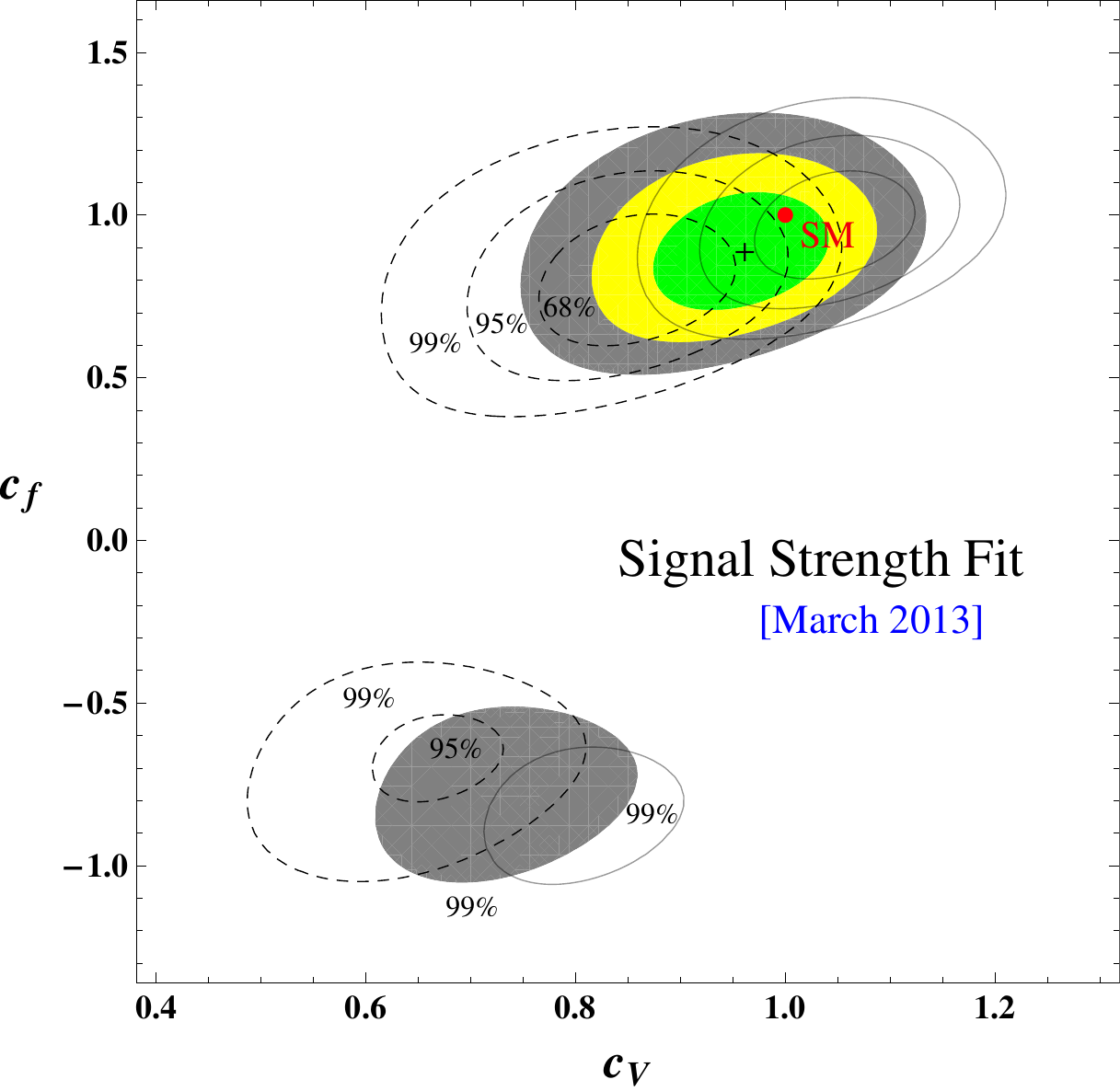}
\hspace{0.7cm}
\includegraphics[width=0.45\textwidth,height=8cm]{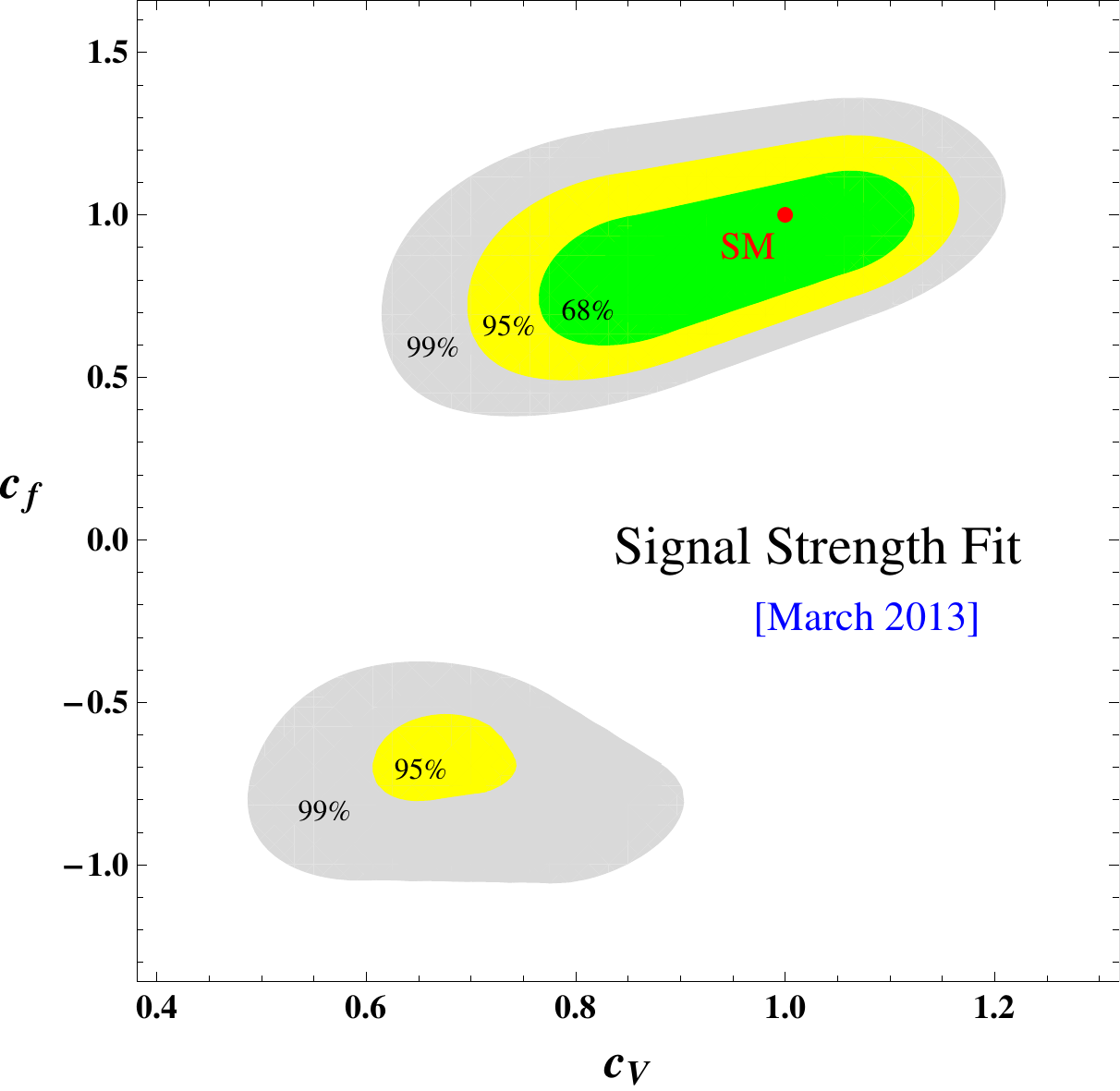}
\end{tabular}
\caption{{\small \underline{Left plot} Best-fit regions at $68.27\%{\rm CL}$ (green), $95.45\%{\rm
CL}$ (yellow) and $99.73\%{\rm CL}$ (grey) in the plane $c_f$ versus $c_V$,
based on the $\chi^2$ function of eq.~(\ref{eq:Chi2def}); the best-fit location
is indicated by a (black) cross. The `concentric' best-fit domains at the same
CLs obtained also from $\chi^2$ but for the two extreme theoretical predictions
-- upper (plain contours) and lower (dashed contours) -- of the Higgs signal strengths, 
are presented. The SM (red) point at $c_f=c_V=1$ is also shown.
\underline{Right plot} Best-fit domains at the $68.27\%{\rm CL}$ (green), $95.45\%{\rm
CL}$ (yellow) and $99.73\%{\rm CL}$ (light grey) based on the $\chi^2$ function; these domains
were obtained by varying continuously the Higgs signal strengths from their lowest to highest theoretical predictions.}}
\label{fig:mu}
\end{center}
\end{figure}

The results of the fit of the data is presented in Fig.~\ref{fig:mu} and  relies
on established values~\cite{PDG} of  $\Delta \chi^2 = \chi^2 - \chi^2_{\rm
min}$ with the following $\chi^2$ function 
\begin{eqnarray}
\chi^2 \ = \ \sum_{i} \frac{[\mu_{i}(c_f,c_V)-\mu_{i}\vert_{\rm exp}]^2}{(\delta\mu_{i})^2} \ .
\label{eq:Chi2def}
\end{eqnarray}
In addition to the fit that is usually performed, leading to the colored regions on the left plot 
of Fig.~\ref{fig:mu} for the best fits at the $1\sigma, 2\sigma$ and $3\sigma$
levels in the $c_f$ versus $c_V$ plane~{$^8$}\footnotetext[8]{Because of the exact reflection
symmetry under, $c \to -c$, leaving the squared amplitudes of the Higgs rates
unaffected, only the part, $c_V>0$, is presented in Fig.~\ref{fig:mu}.}, we also
present the results of the fit for a treatment of the theoretical uncertainty as
a bias.  The value, $\mu_{i} \vert_{ \rm exp} +
(\delta \mu_i/\mu_i \vert_{\rm th}) \Big\vert \mu_{i}\vert_{\rm exp} \Big\vert$, corresponds to the plain contours on Fig.~\ref{fig:mu} while the dashed
contours are for the lower $\mu_{i}\vert_{\rm exp}$ expectations
(negative sign in front of the error).  The distances
between the two contours represent the theoretical uncertainty induced on the
fitted parameters. The treatment of the theoretical uncertainty as
a bias is also illustrated on the right plot of Fig.~\ref{fig:mu} using colored domains. 

A first conclusion that can be drawn from this figure is that, with the latest
LHC data, in particular  the new CMS di-photon rate which has no excess compared
to (and is even slightly below) the SM expectation, the SM point is now included
inside the $1\sigma$ domain. Before the CMS  update, there was a large
deviation  in the di--photon channel which triggered discussions about   the
possibility that $c_f<0$ (as shown in the figure for the $2\sigma$ and $3\sigma$
regions) which leads to a constructive interference between the top quark and
$W$-boson loop contributions to increase the di-photon rate. This possibility has
been elaborated within several effective scenarios in the recent
literature to generate  such a di--photon enhancement (see examples in
Ref.~\cite{Fits}).

The statement that the fit result is compatible  with the SM expectation  is
true regardless of  the method chosen to implement the theoretical uncertainty.
Nevertheless, it appears clearly on the figure that treating this uncertainty 
as a bias enhances significantly its effects  in the $c$-parameter
determination. For instance, the entire $1\sigma$ domain on the plot is larger
when taking into account the possible shifts due to the  theoretical
uncertainty. 

As this theoretical uncertainty cancels out in the ratios $D_{XX}$ of signal
strengths of  eqs.~(\ref{Dapprox},\ref{ratio1}), we have performed a fit based
on the $\chi_R^2$  function~{$^9$}\footnotetext[9]{The fact that the distribution of $\mu$
ratios is not gaussian is not expected to modify our  results. Also,
we refrain from including the $H\to b\bar b$ channel as the ATLAS and CMS 
sensitivities are still too low~\cite{CONF-2012-170,PAS-HIG-12-045}.}:  
\begin{eqnarray}
\chi^2_R  \  =  \ Ê\frac{[D_{\gamma \gamma}^{gg}(c_f,c_V)-\frac{\mu_{\gamma\gamma}}{\mu_{ZZ}}\vert^{gg}_{\rm exp}]^2}
{\big [ \delta(\frac{\mu_{\gamma\gamma}}{\mu_{ZZ}})_{gg}\big ]^2}  + 
\frac{[D_{\tau \tau}^{gg}(c_f,c_V)-\frac{\mu_{\tau\tau}}{\mu_{WW}}\vert^{gg}_{\rm exp}]^2}
{\big [ \delta(\frac{\mu_{\tau\tau}}{\mu_{WW}})_{gg}\big ]^2}  + 
\frac{[D_{\tau \tau}^{\rm VBF}(c_f,c_V)-\frac{\mu_{\tau\tau}}{\mu_{WW}}\vert^{\rm VBF}_{\rm exp}]^2}
{\big [ \delta(\frac{\mu_{\tau\tau}}{\mu_{WW}})_{\rm VBF}\big ]^2}  .
\label{eq:Chi2R}
\end{eqnarray} 
We have considered the inclusive di-photon channels of CMS~\cite{PAS-HIG-13-001,Ochando} and 
ATLAS~\cite{CONF-2013-012} that are largely dominated the ggF mechanism. 
Regarding the $ZZ$ final state in ATLAS~\cite{CONF-2013-013} and
CMS~\cite{PAS-HIG-13-002}, we have also used inclusive production~{$^{10}$}\footnotetext[10]{In
fact,  for $\mu_{\gamma\gamma}/\mu_{ZZ} \vert_{\rm exp}$, we will use  the more
accurate $\rho_{\gamma\gamma/ZZ}$ value provided by the ATLAS
collaboration~\cite{CONF-2013-034}; this quantity corresponds exactly to the
branching fraction ratio and it is deduced from a combination of the ggF and VBF
channels.}. Finally,  for the $WW$ and $\tau\tau$ searches in
ATLAS~\cite{CONF-2013-030,CONF-2012-160} and in
CMS~\cite{PAS-HIG-13-003,PAS-HIG-13-004},  we have selected Higgs production in
ggF with an associated 0/1 jet or the VBF production mechanism. Hence for both
ATLAS and CMS, the situation is equivalent in a good approximation to have
vanishing selection efficiencies except,  $\epsilon_{gg}^Z \simeq
\epsilon_{gg}^\gamma \simeq 1$ and $\epsilon_{gg}^\tau \simeq \epsilon_{gg}^W
\simeq 1$ or $\epsilon_{\rm VBF}^\tau \simeq \epsilon_{\rm VBF}^W \simeq 1$, so
that the theoretical predictions for the ratios simply read as in
eq.~(\ref{ratio1}).  

\begin{figure}[t]
\begin{center}
\includegraphics[width=0.45\textwidth,height=6.8cm]{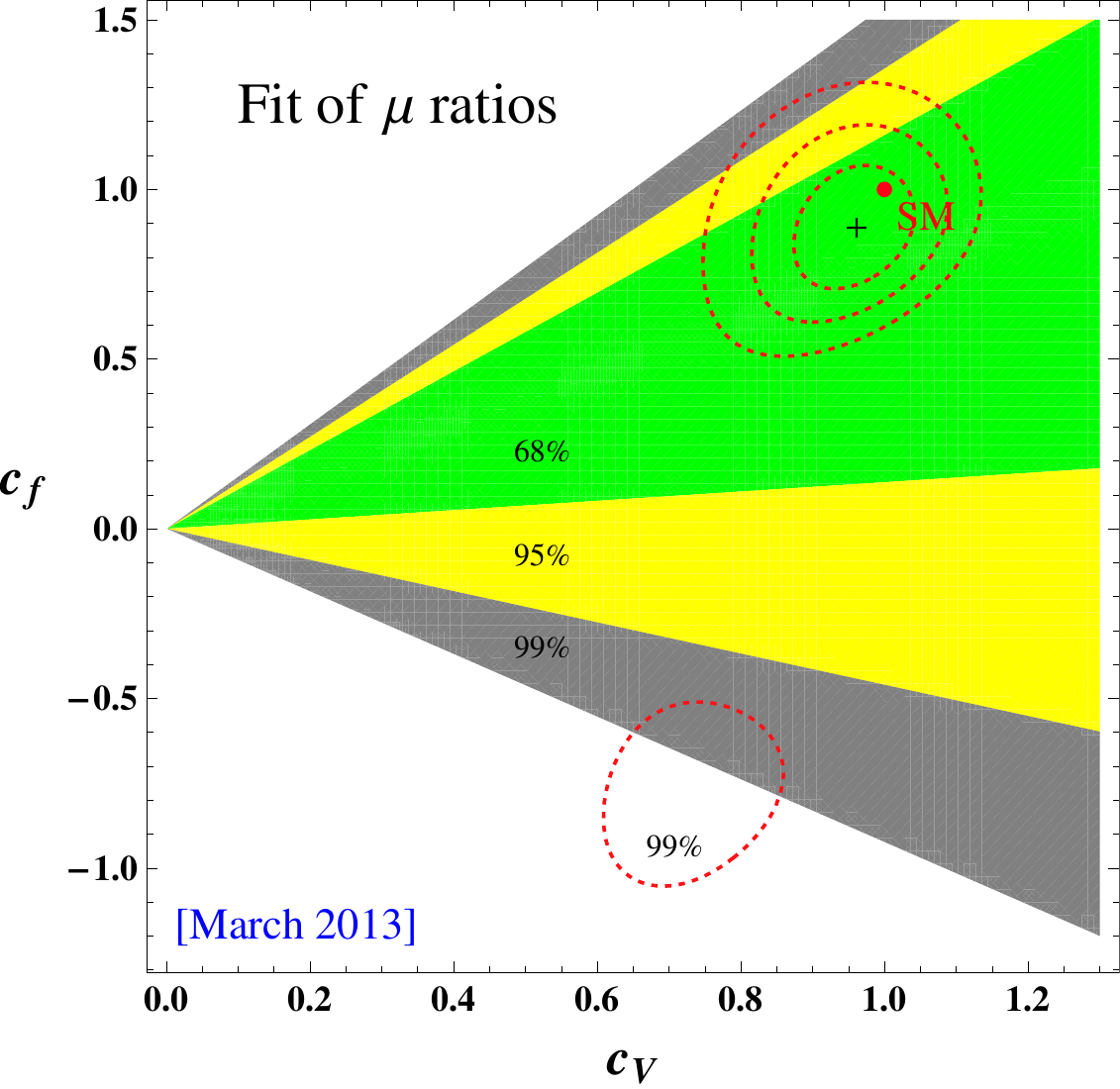}
\hspace{0.7cm}
\includegraphics[width=0.45\textwidth,height=6.8cm]{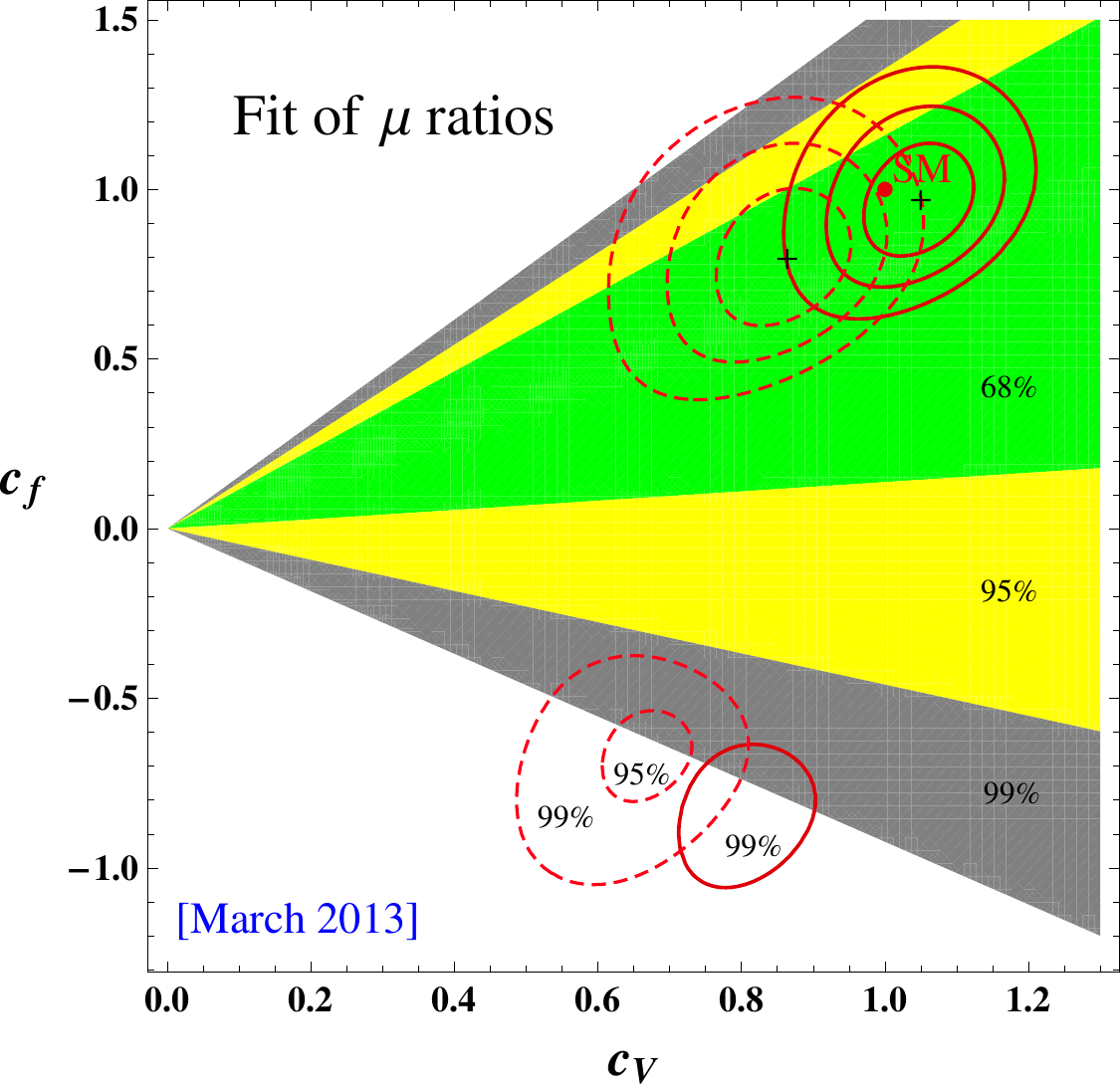}
\caption{\small 
\underline{Left:} Best-fit regions at $68.27\%{\rm CL}$ (green), $95.45\%{\rm
CL}$ (yellow) and $99.73\%{\rm CL}$ (grey) in the plane $c_f$ versus $c_V$,
based on the $\chi_R^2$ function.   The best-fit (dotted) contours obtained from
the $\chi^2$ function in case of a theoretical error added in quadrature (as in
Fig.~\ref{fig:mu})  are superimposed (in red). The associated best-fit point
(cross) and SM (red) point are also shown. \underline{Right:} Same plot as the
left one but the best-fit domains from the $\chi^2$ analysis    are now derived
for the two extreme theoretical predictions of the signal strengths (as in
Fig.~\ref{fig:mu}). \label{fig:ratio}}
\end{center}
\end{figure}

The combined ratio values measured by the ATLAS and CMS collaborations are
\beq 
\frac{\mu_{\gamma\gamma}}{\mu_{ZZ}} \Big\vert_{\rm exp}=1.1^{+0.4}_{-0.3}~, \ \  
\frac{\mu_{\tau\tau}}{\mu_{WW}}\Big\vert^{\rm VBF}_{\rm exp}= - 0.24 \pm 0.83 \ \ {\rm and } \ \  
\frac{\mu_{\tau\tau}}{\mu_{WW}}\Big\vert^{gg}_{\rm exp} =1.2 \pm 0.75
\eeq
The errors  $\delta(\mu_{\gamma\gamma}/{\mu_{ZZ}})$ and  $\delta({\mu_{\tau\tau
}} / {\mu_{WW}})$ are computed assuming no correlations between the different
final state searches.  These uncertainties on the ratios are derived from the
individual errors, $\delta\mu_i$ -- provided in the experimental papers --  and
are thus also dominated by the experimental uncertainties, e.g. 
 $\delta( {\mu_{\gamma\gamma}}/{\mu_{ZZ}}) \approx
\delta({\mu_{\gamma\gamma}}/{\mu_{ZZ}})\vert_{\rm exp}$, as expected from  
the fact that the theoretical uncertainty largely cancels in  ratios
$D_{\gamma\gamma}$ and $D_{\tau \tau}$. These ratios are given by 
{\small \beq D_{\gamma \gamma} &
\simeq &  \frac{1}{\vert c_Z \vert^2} \ \bigg \{ \frac{\big\vert \frac{1}{4}
c_W A_1[m_W] + (\frac{2}{3})^2 c_t   A[m_t] + (-\frac{1}{3})^2 c_b A[m_b] +
(\frac{2}{3})^2 c_c A[m_c] + \frac{1}{3} c_\tau A[m_\tau] \big\vert^2} 
{\big\vert \frac{1}{4} A_1[m_W] + (\frac{2}{3})^2  A[m_t] + (-\frac{1}{3})^2 
A[m_b] + (\frac{2}{3})^2 A[m_c] + \frac{1}{3} A[m_\tau] \big\vert^2} \bigg \} 
\nonumber
\\ D_{\tau \tau} & \simeq &  \frac{\vert c_\tau \vert^2}{\vert c_W \vert^2} \ ,
\label{Eq:final} \eeq } 
where  $A[m] \equiv A_{1/2}[\tau(m)]$ and $A_1[\tau(m)]$ are
respectively the form factors for spin~1/2 and spin~1 particles~\cite{Review} 
normalized such that $A[\tau(m)\ll 1]\to 1$ and $A_1[\tau(m)\ll 1]\to -7$ with
$\tau(m)=M_H^2/4m^2$ and, for $m_H \simeq 125$~GeV,  one has
$A_1[\tau(m_W)] \simeq -8.3$ and $A_{1/2}[\tau(m_t)] \approx 1$.

In Figure~\ref{fig:ratio}, we show the results from fitting the Higgs decay
ratios through the function $\chi_R^2$. In the left panel,  the best-fit domains
obtained e.g. at $1\sigma$ do not exclude parts of the $1\sigma$ regions
obtained from  $\chi^2$; such a compatibility was expected since the main
theoretical uncertainty cancels out in the $D_{XX}$ ratios   and is negligible
for the signal strengths since it is added in quadrature to the experimental
error as already described.  The domains from $\chi^2$ are even more restricted
as ${\it (i)}$ this function exploits the full experimental  information on the
Higgs rates and not only on the ratios and ${\it (ii)}$ the experimental  error
on a ratio of rates is obviously higher than on the rates alone.

In the case where the theoretical error for each Higgs channel is taken
into account as a bias, the best-fit contours span wider regions of the
parameter space. This could be seen  in Fig.~\ref{fig:mu} and the same contours
appear  in the left--hand side of Fig.~\ref{fig:ratio} as well as in its 
right--hand side part where the theoretical uncertainty enters as a bias for
$\chi^2$. In the latter case, the large $1\sigma$ domain from $\chi_R^2$
excludes a small part of the $1\sigma$ region (near the SM point) obtained from
$\chi^2$. Besides, the lower $95.45\%{\rm CL}$ region from the $\chi^2$ fit is
completely excluded by the $\chi_R^2$ domain (in yellow) at the same ${\rm CL}$.
In conclusion, within the more realistic case of treating the theoretical 
uncertainty as a bias, the $\chi_R^2$ domains can thus play an important role by
excluding parts of the $\chi^2$-fit regions;  this is due to the increased
contribution of the theoretical error, in $\chi^2$, which does not affect
$\chi_R^2$.

\begin{figure}[t]
\begin{center}
\includegraphics[width=0.31\textwidth,height=5cm]{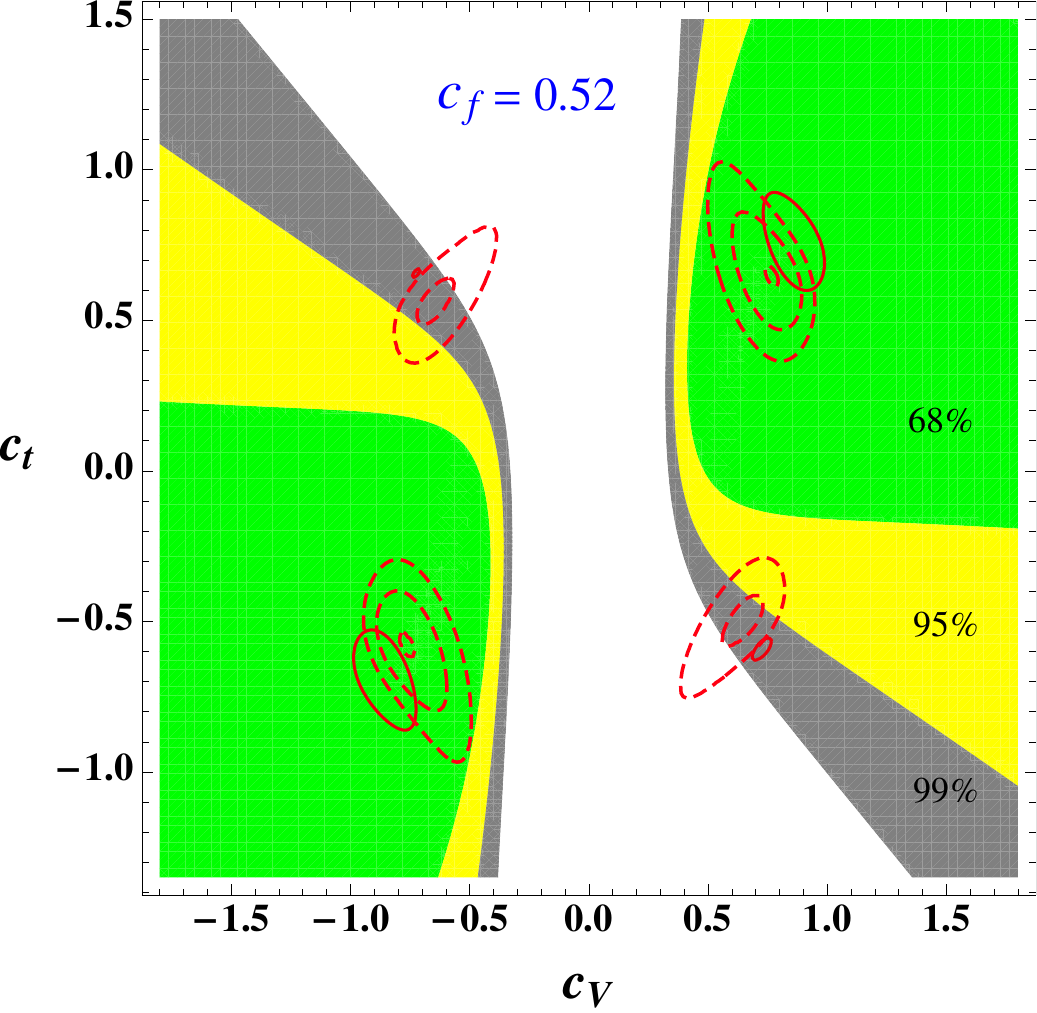}
\hspace{0.2cm}
\includegraphics[width=0.31\textwidth,height=5cm]{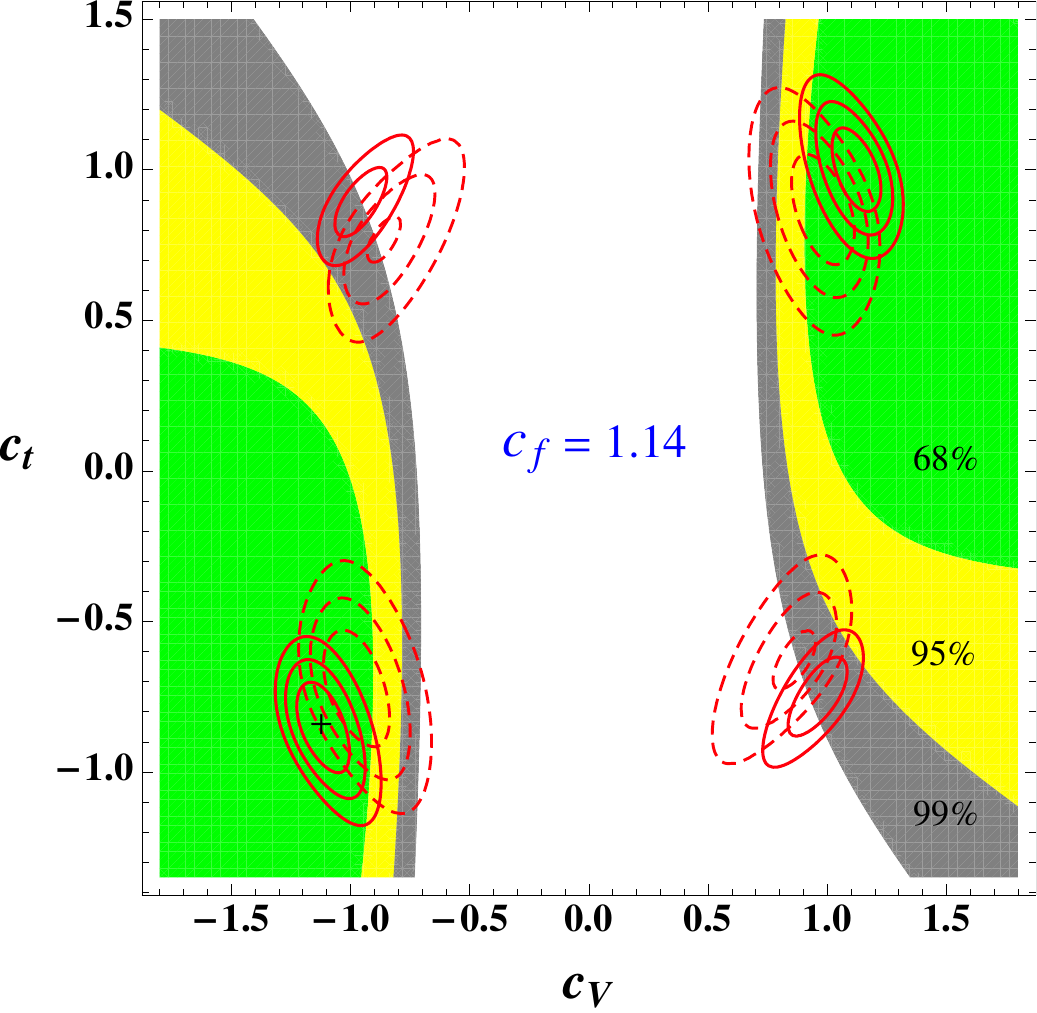}
\hspace{0.2cm}
\includegraphics[width=0.31\textwidth,height=5cm]{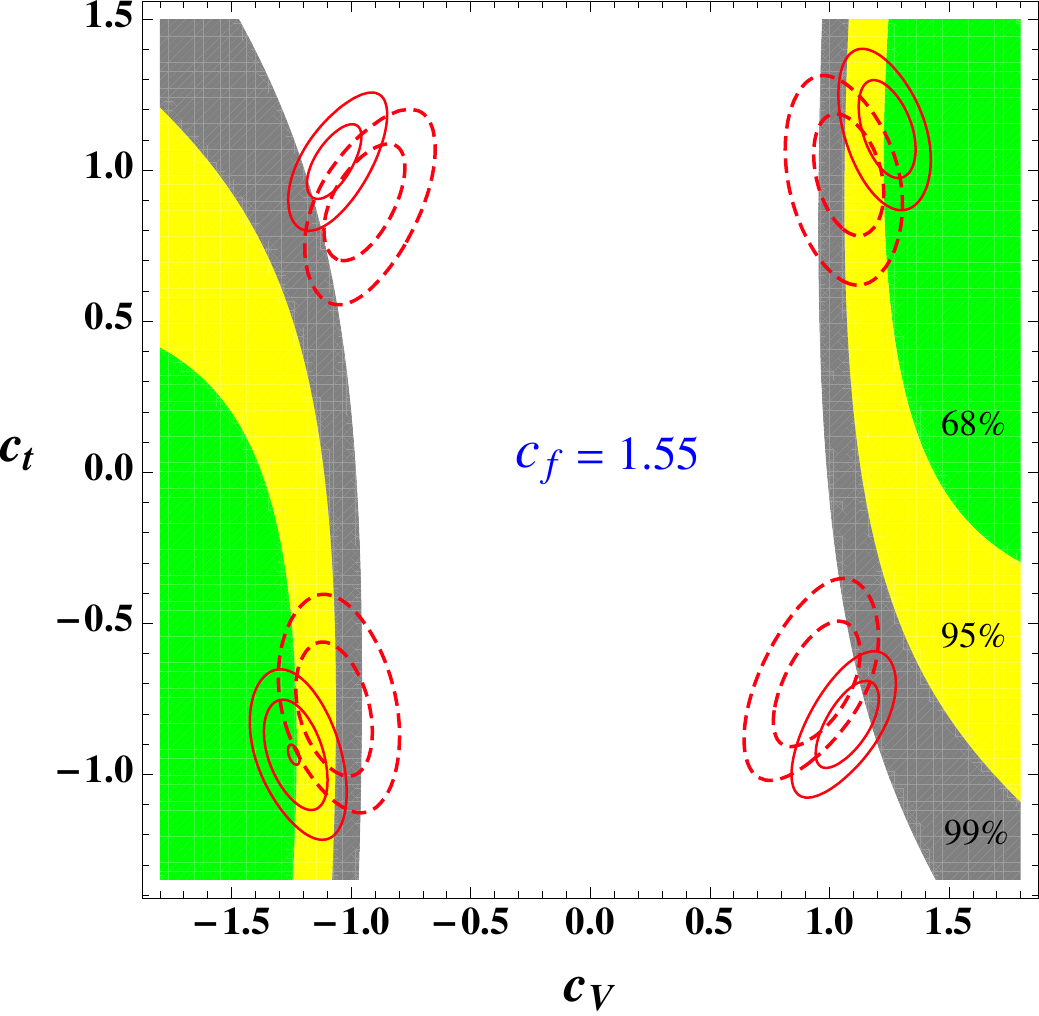}
\caption{\small 
Best-fit regions at $68.27\%{\rm CL}$ (green), $95.45\%{\rm CL}$ (yellow) and
$99.73\%{\rm CL}$ (grey) in the plane $c_t$ versus $c_V$ as obtained from a
three-dimensional fit  (whose best-fit point is the black cross on the central
plot) of the 3 free parameters, $c_f$, $c_t$, $c_V$,   based on the $\chi^2_R$
function. The three two-dimensional plots correspond to the slices of these
three-dimensional domains at $c_f=0.52$,  $1.14$ and $1.55$. Superimposed are
the best-fit domains at $68.27\%{\rm CL}$, $95.45\%{\rm CL}$, $99.73\%{\rm CL}$ 
obtained from $\chi^2$ for the two theoretical signal strength predictions --
upper (red plain) and lower (red dashed contours).  
\label{fig:ctop}}
\end{center}
\end{figure}

In Fig.~\ref{fig:ctop}, we present the three-dimensional $\chi^2$-fit results in
the case where the parameters $c_f$, $c_t$ and $c_V$ are free. It shows how
precisely are presently known the top quark Yukawa and the gauge boson
interactions with the Higgs scalar  (in the  case of a preserved custodial
symmetry). For either lower (left plot) or higher (right plot)  bottom quark
Yukawa couplings as compared to the SM, the size of the characteristic $1\sigma$
regions decrease significantly, which also gives an idea of the present
knowledge of the coupling $c_b$.   This more realistic three-dimensional fit
illustrates as well the potential interest of the $\chi^2_R$-fit~: for instance,
one observes on the central plot, that it excludes  at $1\sigma$ the lower-right
(i.e. the dysfermiophilia solution $c_t<0$) and upper-left (i.e. its almost
symmetric domain) $1\sigma$ regions resulting from the $\chi^2$-fit.

Since $\chi_R^2$ is only affected by the experimental uncertainty, it is
interesting to quantify the evolution of the fit when the experimental 
systematic and statistical errors are reduced~\cite{EurStrat}.  For this
purpose, we combine the present measurements with the expected  results from the
$14$~TeV LHC in each channel investigated by ATLAS and CMS. We assume the
central values at $14$~TeV to be identical to those from the combination of the
$7$ and $8$~TeV data, and that the future experimental errors,
$\delta\mu_i\vert_{\rm exp}$, will reduce  essentially like   the inverse of the
square roots of number of events, $\sqrt{ \sigma_i {\cal L} }$ with ${\cal L} $
the integrated luminosity~{$^{11}$}\footnotetext[11]{This is justified for the  statistical error
and corresponds to an optimistic situation for the systematic error,  which is
difficult to predict for each channel but  depends partially on the background
rate uncertainties which have a statistical behavior as well.}. 

\begin{figure}[t]
\begin{center}
\includegraphics[width=0.47\textwidth,height=7cm]{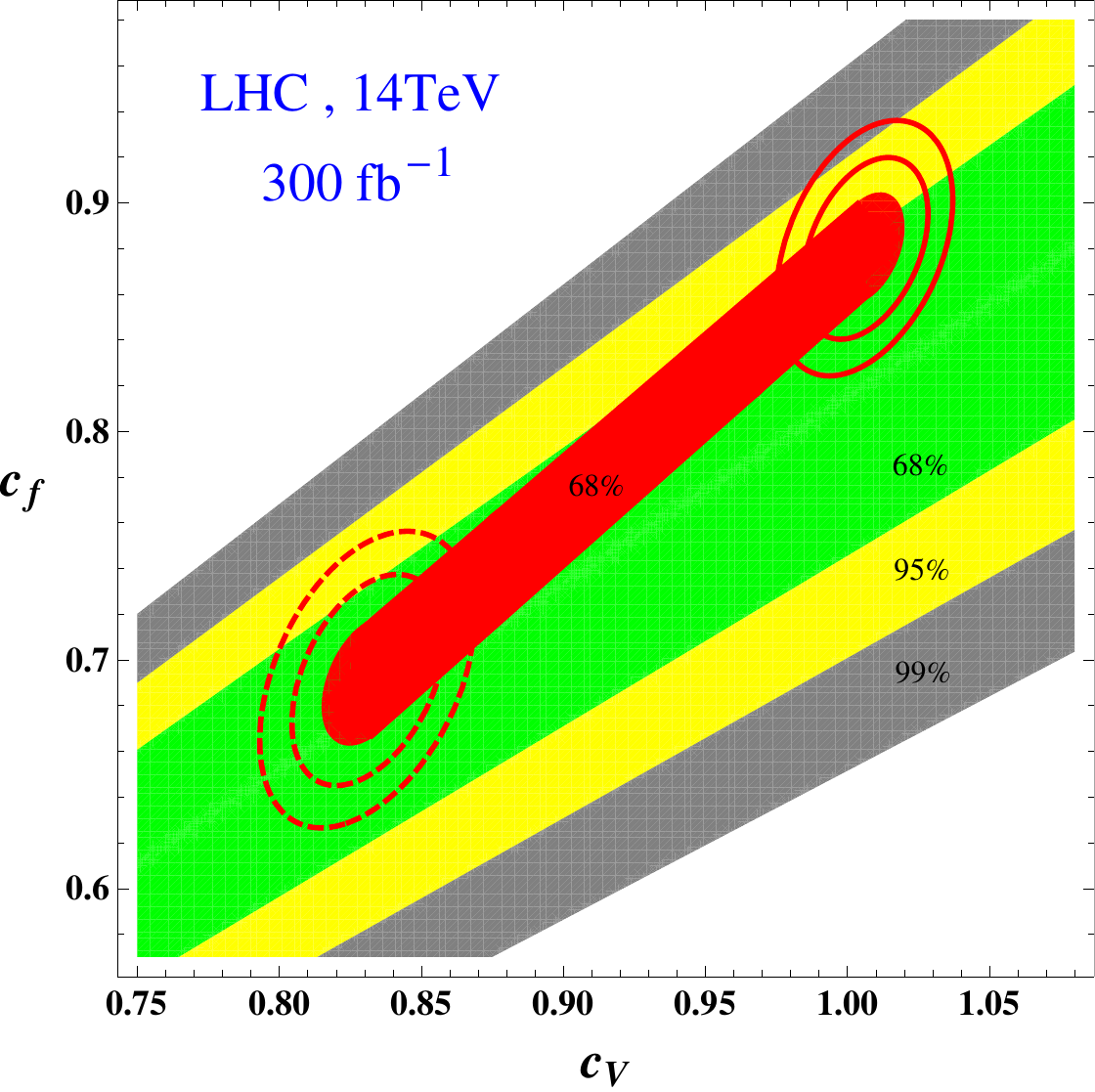}
\hspace{0.4cm}
\includegraphics[width=0.47\textwidth,height=7cm]{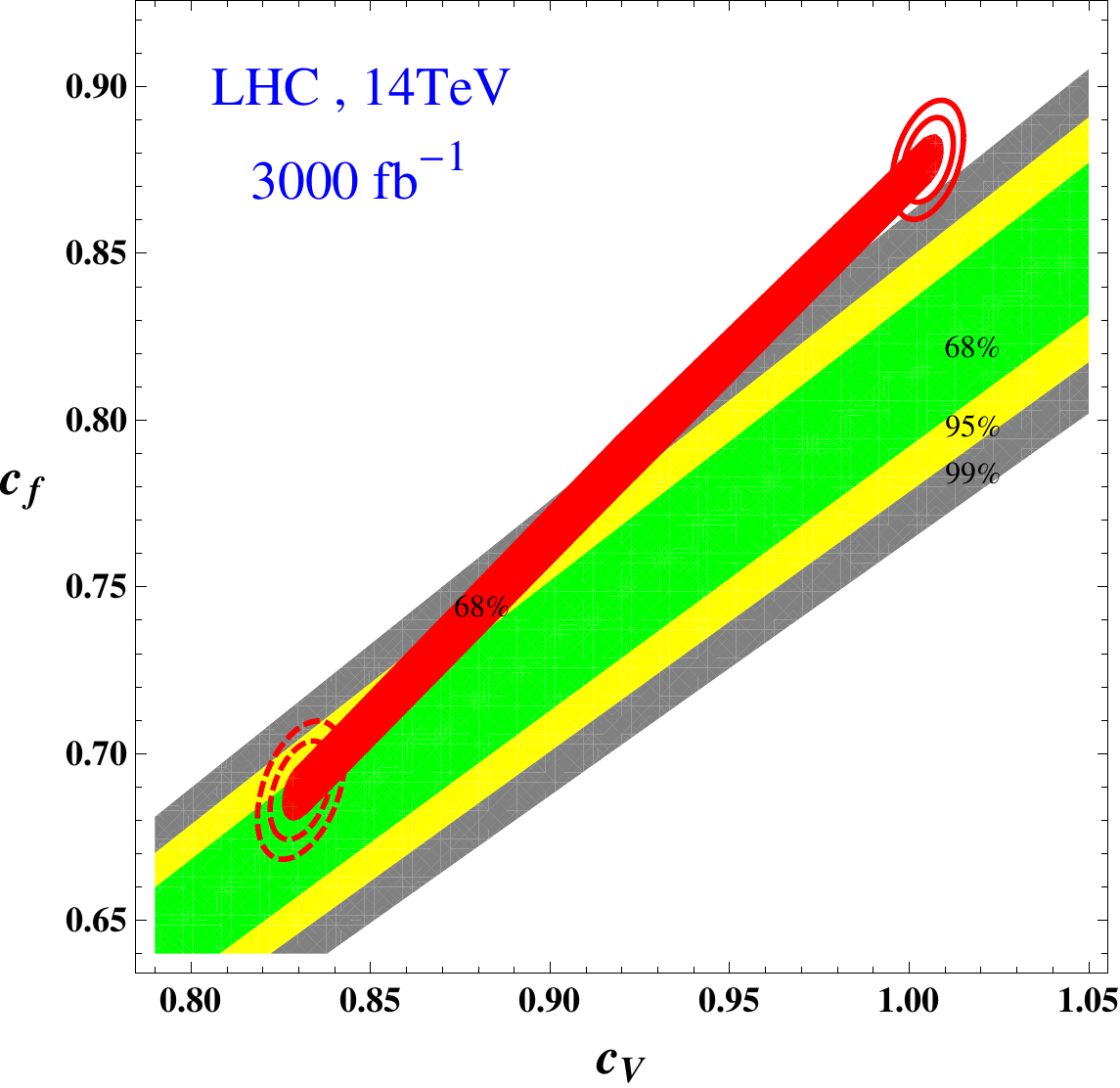}
\caption{\small 
Best-fit regions at $68.27\%{\rm CL}$ (green), $95.45\%{\rm CL}$ (yellow) and
$99.73\%{\rm CL}$ (grey) in the plane $c_f$ versus $c_V$, based on the
$\chi_R^2$ function and including hypothetical data from the $14$~TeV LHC with
${\cal L} =300$~fb$^{-1}$ [left plot] or $3000$~fb$^{-1}$ [right plot]. The 
best-fit $\Delta \chi^2$ contours at $95.45\%{\rm CL}$ and $99.73\%{\rm CL}$ obtained in 
the same conditions, for the two extreme theoretical predictions of signal strengths (red plain and dashed ellipses),  
are superimposed; the $68.27\%{\rm CL}$ domain presented (in red) was obtained by varying continuously the
signal strengths from their lowest to highest theoretical predictions. So typically the length of this domain indicates
the theoretical uncertainty and its width the experimental error. The exactly symmetric domains, obtained via
$c_f \to -c_f$, $c_V \to -c_V$, are not shown. \label{fig:ratioFUT}}
\end{center}
\end{figure}

The estimated $\chi^2$ fit results at $14$~TeV are presented in
Fig.~\ref{fig:ratioFUT} assuming luminosities of ${\cal L} \equiv300$  and
$3000$~fb$^{-1}$~\cite{EurStrat}. The behavior of the best-fit $\chi_R^2$
regions appearing  in the figure originates from the compensation between the
enhancement of $\Gamma ( H\! \to\! \gamma\gamma)$ and that of $\Gamma ( H\! \to
\! ZZ)$ as $c_V$ increases, leading to relatively stable $D_{\gamma \gamma}$
values;  the increase of $\Gamma ( H\! \to\! \tau\tau)$  and $\Gamma ( H\! \to\!
WW)$  with increasing $c_\tau$ or $c_W$ also compensate each other in $D_{\tau
\tau}$.  Best-fit values of $c_f$ and $c_V$ in Fig.~\ref{fig:ratioFUT}
would be illustrative only since the precise central values are
of course not yet known, neither the exact experimental uncertainties. However, the
above estimation of the statistical error provides  an indication of the  
typical relative sizes of the best-fit $\chi^2$ and $\chi_R^2$ domains in the
future.

The main features that the plots of Fig.~\ref{fig:ratioFUT} exhibit are  that
when increasing the luminosity and hence, reducing the experimental error as
shown by the smaller ellipses on the right plot, the $\chi^2$-fit reaches the
level where the theoretical error is dominating and fixes the typical
uncertainty scale  (stable red band length on the two plots),   whereas for the
$\chi_R^2$-fit in which the theoretical uncertainty is absent, the precision
obtained on the couplings $c_f$ and $c_V$ improves  as long as the experimental
error decreases (decrease of the colored region widths on the right plot). 
Thus, for high LHC luminosities, the fit of the decay ratios will play a crucial role and will have to be combined 
with the common Higgs rate fit, as illustrated on the right plot of Fig.~\ref{fig:ratioFUT}: there for instance the $1\sigma$ region from $\chi^2$ 
(typically the red band) is wider than when restricted to its intersection with the $\chi_R^2$ domain at $1\sigma$ (green band).  
This corresponds, to an improvement of the whole accuracy from $\sim 10\%$  
down to $\sim 5\%$ on both $c_f$ and $c_V$; with such accuracies one starts to be really
sensitive to deviations in the Higgs couplings arising in supersymmetric
theories or composite Higgs models as, for instance,  discussed in
Ref.~\cite{Gupta}.

\begin{figure}[t]
\begin{center}
\includegraphics[width=0.47\textwidth,height=7cm]{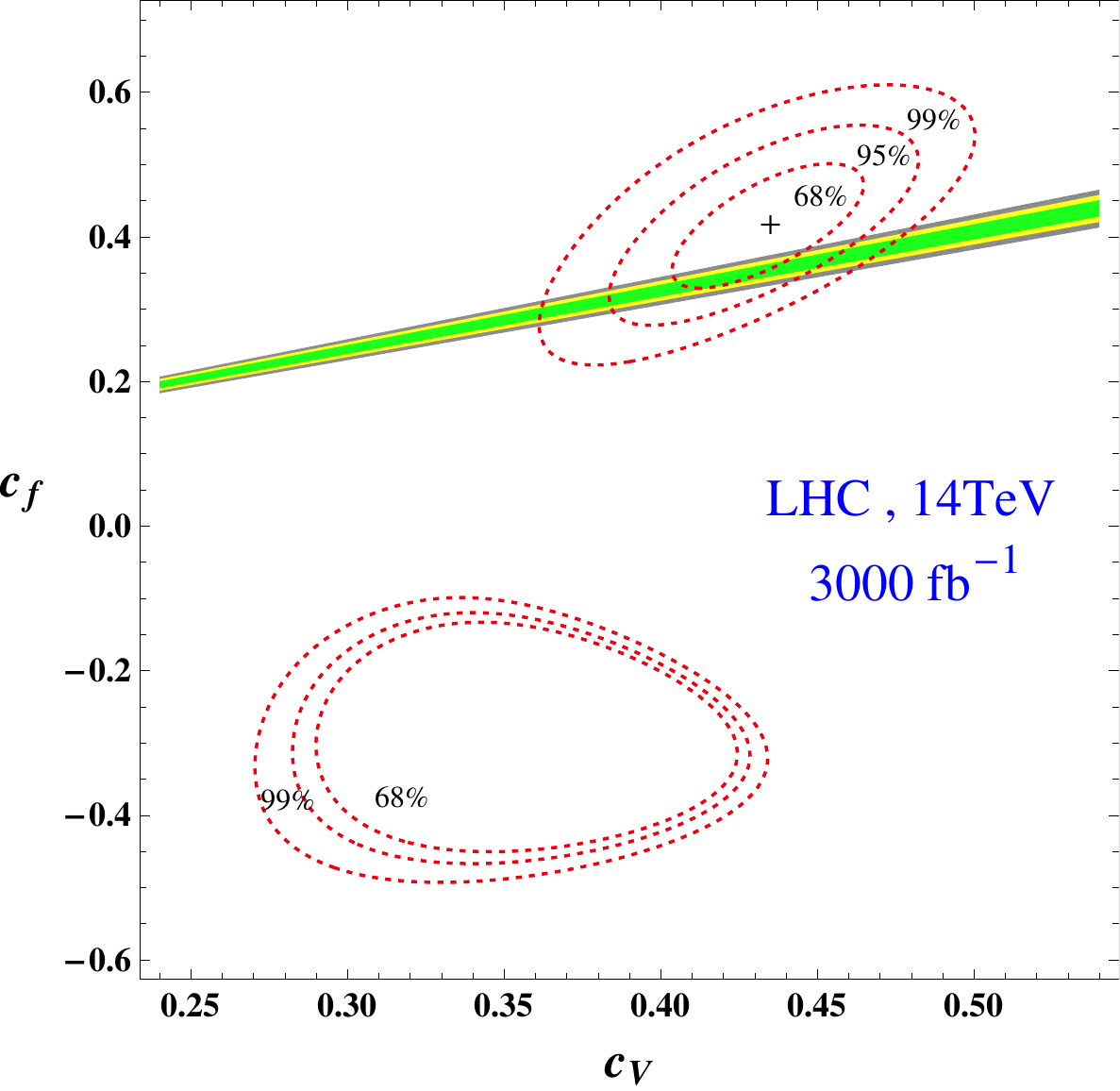}
\caption{\small 
Best-fit regions at $68.27\%{\rm CL}$ (green), $95.45\%{\rm CL}$ (yellow) and
$99.73\%{\rm CL}$ (grey) in the plane $c_f$ versus $c_V$, based on the
$\chi_R^2$ function and including hypothetical data from the $14$~TeV LHC with
${\cal L} =3000$~fb$^{-1}$ [as in Fig.~\ref{fig:ratioFUT}].
The best-fit $\Delta \chi^2$ contours 
at $68.27\%{\rm CL}$, $95.45\%{\rm CL}$, $99.73\%{\rm CL}$ obtained in
the same conditions, and with the theoretical error added in quadrature,
are superimposed as dotted (red) contours; the best central point is indicated
as a (black) cross. 
\label{fig:ratioQUAD14}}
\end{center}
\end{figure}

For illustration, we present in Fig.~\ref{fig:ratioQUAD14} the expected results of the 
$\chi^2$ fit, at $14$~TeV with ${\cal L} \equiv 3000$~fb$^{-1}$, when adding the theoretical error to the experimental one 
in quadrature; the associated best-fit regions are obviously different from the pairs of ellipse-like best-fit domains obtained 
for a theoretical uncertainty treated as a bias [Fig.~\ref{fig:ratioFUT}]. Furthermore, in Fig.~\ref{fig:ratioQUAD14}, there exist  
best-fit regions at negative $c_f$ values. This comparison between Fig.~\ref{fig:ratioFUT} and Fig.~\ref{fig:ratioQUAD14} 
clearly allows to convince oneself that the choice of the treatment of theoretical errors will be crucial for the determination of
the Higgs couplings. Nevertheless, even in the case of a theoretical uncertainty combined in quadrature, the fit of rate ratios 
(independent of the theoretical error and presented again in Fig.~\ref{fig:ratioQUAD14}) allows to select a sub-part of the
$1\sigma$, $2\sigma$, $3\sigma$ ellipses derived from the signal strength fit.

\subsection*{{\Large 4. The parity or CP--composition of the Higgs boson}}

As mentioned in the introduction, the observables such as correlations in Higgs
decays into vector boson pairs~\cite{CPH-d} or in Higgs production with or
through these states~\cite{CPH-p} that are usually used to probe the Higgs
parity project out only the CP--even component of the $HVV$ coupling even if the
state has both CP--even and CP--odd components. Thus, in these CP studies,  one
is simply verifying, a posteriori, that a CP--even Higgs state has been  indeed
produced. The $HVV$ coupling takes the  general form (here, we assume $c_V>0$)
\begin{eqnarray}
g_{HVV}^{\mu \nu} = -i \ c_V (M_V^2/v)\; g^{\mu \nu}
\end{eqnarray}
where $c_V$ measures the departure from the SM: $c_V\!=\!1$ for a pure CP--even 
state with SM--like couplings and $c_V=0$ for a pure CP--odd state. Indeed, the
coupling of a pseudoscalar $A$ state to $W/Z$ bosons is zero at tree-level and
is generated only through loop corrections which are expected to  tiny.  The
measurement of $c_V$ should allow to determine the CP composition of a Higgs
boson if it is indeed a mixture of CP--even and CP--odd states.  

However, having $c_V\! \neq\! 1$ does not automatically imply that the observed
state has a pseudoscalar component. As a matter of fact, the Higgs sector could
be enlarged to contain other  neutral Higgs particles $H_i$ that have not been 
detected so far because they are too heavy or too weakly coupled.  In this
case,  the sum of the squared couplings of each state $H_i$ to gauge bosons,
$c_{Vi}^2\, g_{HVV}^2$,  should reduce to the SM Higgs coupling, $g_{HVV}^2$.
Hence, $c_V^2 \!<\! 1$, could mean  that there are other CP--even states which
share the SM Higgs coupling to $VV$ with the observed Higgs boson. Nevertheless,
in all cases, the quantity  $\kappa_{\rm CP}=1-c_V^2$ gives an {\it upper bound}
on the CP--odd contribution  to the $HVV$ coupling~{$^{12}$}\footnotetext[12]{The best example of
an extended Higgs sector with CP--violation is the minimal supersymmetric
extensions of the Standard Model  (MSSM) with complex soft--SUSY breaking
parameters~\cite{Review2}. One has then three neutral Higgs states $H_1, H_2$
and $H_3$ with indefinite parity and  their CP--even components will share the
SM $HVV$ coupling, $c_{V1}^2+c_{V2}^2+c_{V3}^2=1$. There are no antisymmetric
CP--odd couplings  $H_i V_{\mu \nu} \widetilde{V}^{\mu \nu}$ at tree-level
and those generated at the one--loop level are extremely tiny~\cite{Review2}.}.

In contrast to the couplings to massive gauge boson, the CP--even and CP--odd
components of the state can couple to fermions with the same magnitude and one can write 
\begin{eqnarray}
g_{Hff } = -i \frac{m_f}{v}  \bigg[ {\rm Re}\, (c_f) +i \ {\rm Im}\,( c_f)\,   
\gamma_5 \bigg] \label{CP-Hff}
\end{eqnarray}
where in the SM one has Re$(c_f)\!=\!1$ and Im$(c_f)\!=\!0$ but in general, the 
normalisation of the coupling, ${\rm Re} (c_f)^2 + {\rm Im} (c_f)^2=  |c_f|^2$,
should be taken arbitrary as in the previous section.

Hence, one can consider the same effective Lagrangian as in
eq.~(\ref{Eq:LagEff}) where $c_V=c_Z=c_W$ represents exclusively the CP--even
component of the observed boson assumed to be one eigenstate of an enlarged 
Higgs sector. In contrast, the $c_f\!=\!c_b\!=\!c_c\!=\!c_\tau$ and $c_t$
parameters for light fermions and the top quark contain the CP compositions of 
eq.~(\ref{CP-Hff}) with the possibility of a deviation of the normalisation  
$|c_f|^2$  compared to the SM Yukawa interaction.

In the case of the light fermions, one has $M_H \gg m_f$ so that chiral symmetry
holds and the partial decay widths (the only place where they enter if one
neglects their tiny contribution to the loop induced vertices)  can be simply
written as $\Gamma(H \to f  \bar f) \propto {\rm Re} (c_f)^2 + {\rm Im} (c_f)^2
\propto |c_f|^2$ and the discussion in the previous section should entirely
hold.  

In the case of the top quark, the situation is  different as $m_t\!>\!M_H$.
 The top quark enters the   $Hgg$ and $H\gamma\gamma$ vertices and the
loop form factors for the CP--even $A_{1/2}^H$ and CP--odd $A_{1/2}^A$  parts
are in principle different~\cite{Review}. Fortunately, in these vertices the 
approximation $m_t \gg M_H$ is extremely good and in this limit, the form
factors take simple forms: $A_{1/2}^H = \frac43$ and $A_{1/2}^A=2$.
Ignoring the small contributions of the light fermions for simplicity, the Higgs rates normalized to 
the SM expectations can be written as, 
\begin{small}
\beq 
\frac{\Gamma( H \to \gamma\gamma)}{\Gamma( H \to \gamma\gamma)\vert_{\rm SM}} 
&  \simeq & 
\frac{\big \vert \frac{1}{4} c_W A_1[m_W] + (\frac{2}{3})^2 {\rm Re}(c_t)  \big \vert^2
+ \vert (\frac{2}{3})^2 \frac{3}{2} {\rm Im}(c_t) \vert^2} 
{\big \vert \frac{1}{4} A_1[m_W] + (\frac{2}{3})^2  \big \vert^2 }
\nonumber \\ 
\frac{\sigma( gg \to H)}{\sigma( gg \to H)\vert_{\rm SM}} & = & 
\frac{\Gamma( H \to gg)}{\Gamma( H \to gg)\vert_{\rm SM}}  \simeq 
\big \vert {\rm Re}(c_t)  \big \vert^2 + \vert \frac{3}{2} {\rm Im}(c_t)  
\big \vert^2
\label{Eq:widthsCP} 
\eeq 
\end{small}
with $A_1[m_W] \simeq -8.3$ for $M_H \approx 125$ GeV.  For a pure pseudoscalar
state, Re$(c_t)=0$,  there is no $W$ contribution  to the $H\to \gamma \gamma$
rate; there are also no $H\to ZZ$ and $WW$ decays, a possibility that  is 
clearly excluded by the present data as the $4\ell$ and $2\ell 2\nu$ signals  
have been  observed. To quantify the degree of exclusion of this possibility,
one needs to measure $\kappa_{\rm CP}=1-c_V^2$ (and ideally, independently of 
the fermion couplings $c_f$).

\begin{figure}[t]
\begin{center}
\includegraphics[width=0.5\textwidth,height=7cm]{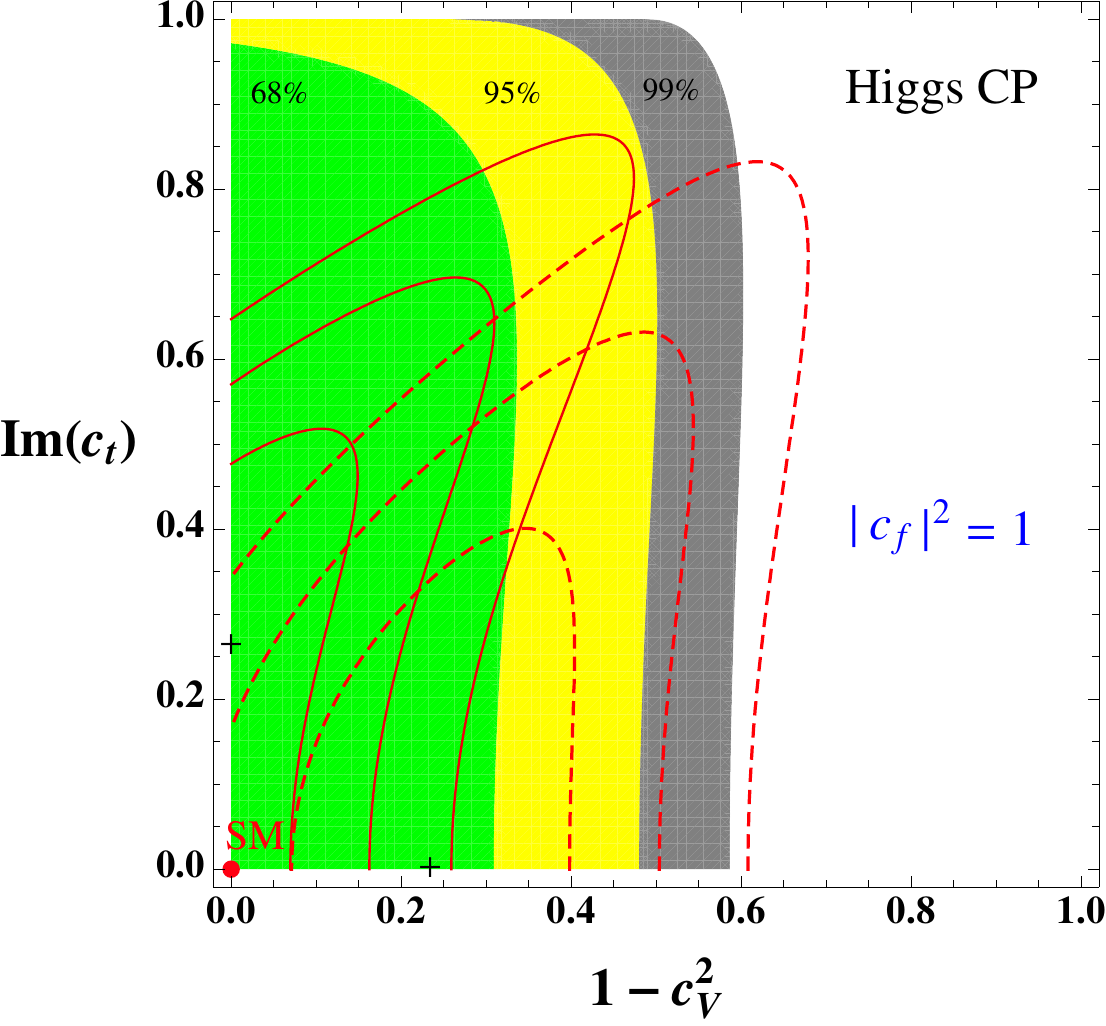}
\caption{\small 
Best-fit regions at $68.27\%{\rm CL}$ (green), $95.45\%{\rm CL}$ (yellow) and $99.73\%{\rm CL}$ (grey) in the plane $1-c^2_V$
versus Im$(c_t)$ for $\vert c_t \vert^2\!=\!\vert c_f \vert^2\!=\!1$~; 
these regions are obtained from a two-dimensional fit based on the $\chi^2_R$ function. 
Superimposed are the best-fit regions at $68.27\%{\rm CL}$, $95.45\%{\rm CL}$, $99.73\%{\rm CL}$ obtained from $\chi^2$ 
for the two theoretical signal strength predictions (plain and dashed contours in red). 
The SM (red) point is represented at the origin together with the best-fit 
points (the two black crosses) for the $\chi^2$ fits. 
\label{fig:CPfit}}
\end{center}
\end{figure}

Based on these rates, one can perform a fit using the same $\chi^2$ function as
in eq.~(\ref{eq:Chi2def}) but with the dependence, $\chi^2 = \chi^2 [\mbox{\rm
Re}(c_t),\mbox{\rm Im}(c_t),c_f,c_V]$. The numerical results are displayed in
Fig.~\ref{fig:CPfit} for the $\chi^2$-fit and the $\chi^2_R$-fit which reveals
itself to be useful as well for measuring the CP-odd component of the Higgs
boson. In this case, we have made the simplifying assumption (besides
$c_V\geq0$) that the absolute normalisation of the fermion couplings is the same
as in the SM,  $\vert c_t \vert^2\!=\!\vert c_f \vert^2\!=\!1$, but in the case
of the top quark, $\mbox{\rm Im}(c_t)$ is assumed to be free. We assume
${\rm Re}(c_t)\geq 0$ and the obtained plot is symmetric under ${\rm Im}(c_t)
\to -{\rm Im}(c_t)$.

The conclusion is that, at the 99.73\%CL or at the $3\sigma$ level,  the CP--odd
component of the observed Higgs boson obeys the upper bound  $\kappa_{\rm
CP}=1-c_V^2 < 0.68$. A pure CP--odd Higgs state, i.e. the case $\kappa_{\rm CP}
\approx 1$, is excluded with more than $4 \sigma$. This is much more
severe than the constraint from the correlations in $H\to ZZ$ decays, which 
(with its inherent limitation discussed above) allows only a $\lsim 3\sigma$ 
discrimination between the CP--even and CP--odd cases 
\cite{CONF-2013-034,PAS-HIG-12-045}.

\subsection*{{\Large 5. The invisible Higgs decay width}}

\begin{figure}[t]
\begin{center}
\includegraphics[width=0.42\textwidth,height=6.9cm]{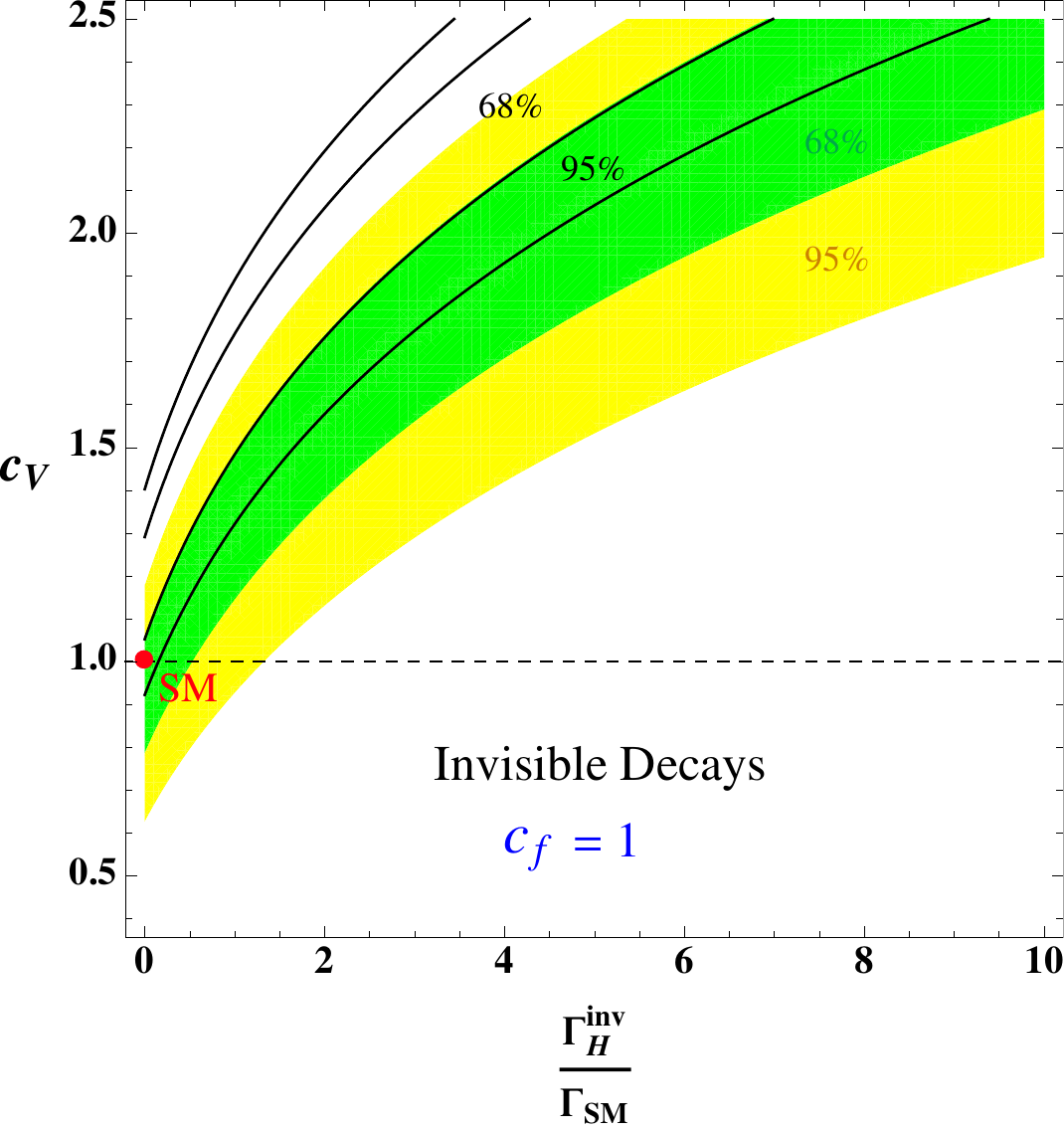}
\hspace{0.6cm}
\includegraphics[width=0.42\textwidth,height=6.9cm]{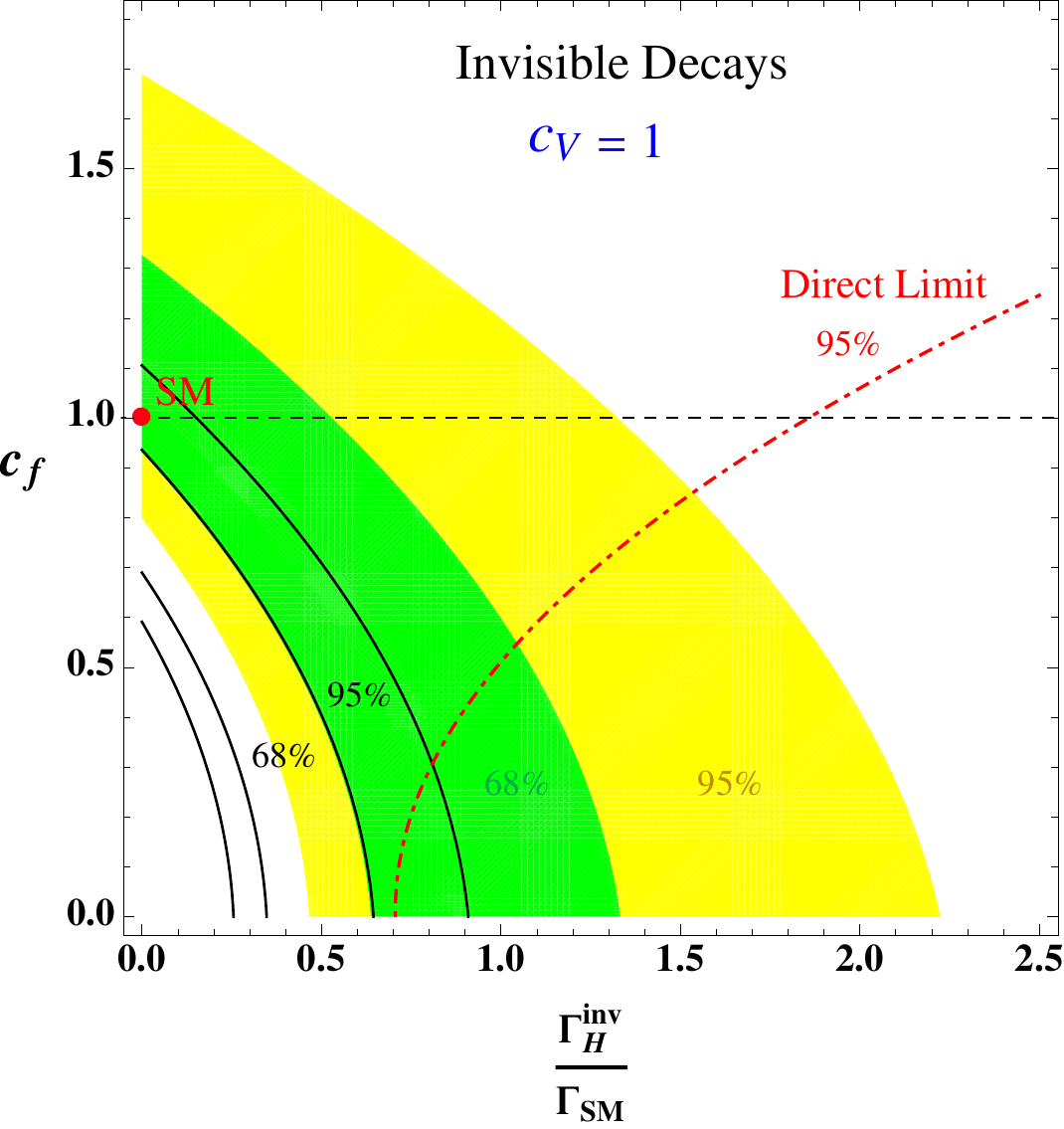}
\caption{\small 
Domains at $1\sigma$ (green), $2\sigma$ (yellow) from the central value of $\mu_{ZZ}$, in the plane $c_V$ (left plot) or $c_f$ (right plot)  
versus $\Gamma_H^{\rm inv} /  \Gamma_H^{\rm SM}$ [$\Gamma_H^{\rm SM}$ being the total SM Higgs width] for $c_t\!=\!1$ and $c_f\!=\!1$ 
(left plot) or $c_V\!=\!1$ (right plot). The dependence of these constraints
on the theoretical uncertainties is illustrated by the (black) curves which 
indicate the other possible extreme domains. The direct upper limit on 
$\Gamma_H^{\rm inv}$ from invisible searches at LHC (derived for 
$c_V\!=\!1$)~\cite{InvATLAS,InvMoriond} is shown on the right plot.
\label{fig:Invfit}}
\end{center}
\end{figure}

In the previous discussion, the signal strength $\mu_{ZZ}$  in the channel $H\to
ZZ\to 4\ell^\pm$ played a prominent role because the theoretical ambiguities 
are minimised: the measurement is inclusive and does not involve the  additional
theoretical uncertainties that are introduced when breaking the cross section
into jet categories and there is no loop induced new physics effect as in the
$H\to \gamma \gamma$ case;  besides that,  it is the most accurate single signal
strength measurement.  One can also use $\mu_{ZZ}$ for the determination of  the
invisible Higgs decay width which enters in the signal strength through the
total decay width $\Gamma_H^{\rm tot}$, $\mu_{ZZ} \propto 
\Gamma (H\to ZZ)/\Gamma_H^{\rm tot}$ with  
\beq 
\Gamma_H^{\rm tot} = \Gamma_H^{\rm inv}+ \Gamma_H^{\rm SM} (c_f,c_V)
\eeq
$\Gamma_H^{\rm SM}(c_f,c_V)$ is the SM total width which is calculated  with
free coefficients $c_f$ and $c_V$ and  including the  state-of-the art radiative
corrections~\cite{BD,BRpaper}.  One can write the  $ZZ$ signal strength as  a
function of $\Gamma_H^{\rm inv}$ and the Higgs  couplings, $\mu_{ZZ}\vert_{\rm
th}  = \mu_{ZZ}(\Gamma_H^{\rm inv},c_V,c_t,c_f =c_c=c_b=c_\tau)$, and impose
that it  lies within its  $1\sigma$  or $2\sigma$ ranges. This  restricts the
parameter space to specific regions  as is shown in Fig.~\ref{fig:Invfit} where
in the left-hand hand side $c_f$ is SM--like and $c_V$ is varied and in the
right--hand side, it is the opposite $c_f$ is varied while $c_V=1$.

On the figure, we also display for comparison the recent direct 
limit~\cite{InvATLAS,InvMoriond} on the invisible Higgs width obtained from combining the 7+8
TeV LHC data in the $q \bar q \to ZH \to Z+\not \hspace{-3mm}E_T$ direct search
channel.  This gives $B_H^{\rm inv}<65\%$ at the $95.45\%{\rm CL}$ if the
assumption $c_V=1$ is made. With the simplifying $c_f=c_V=1$ assumption, the
indirect limit on the invisible width that one obtains from the signal
strengths is better, as can be seen from the figure; at $1\sigma$, it reads as 
\beq
\mu_{\rm inv} = \Gamma_H^{\rm inv}/  \Gamma_H^{\rm SM} \leq 0.52 \ @68\%{\rm CL}
\eeq
This limit is at least a factor of two worse than those obtained in the similar 
fits of Refs.~\cite{Fit-last} using the latest LHC data, the reason being that,
here, we assume a 20\% theoretical uncertainty on the Higgs production cross  
sections that we treat as a bias and do not combine quadratically  with the
experimental  uncertainty.

\subsection*{{\Large 6. Conclusion}}

We have analyzed the  Higgs production cross sections  at the LHC for the
different Higgs decay channels  that have been searched for,  $H\to
WW,ZZ,\tau\tau, b \bar b$ and $H\to  \gamma \gamma$. Using the latest results
given by  the ATLAS and CMS  collaborations with the $\approx 25$ fb$^{-1}$
data collected in the runs at $\sqrt s=7$ and 8 TeV, we have first performed a
fit of the Higgs couplings to fermions and massive gauge bosons and shown that
they are now  compatible with the SM expectation at the $1\sigma$ level. The
accuracy of the various experimental measurements is now almost saturated by
the theoretical uncertainties stemming from QCD.

We have argued that ratio of cross sections times branching ratios  in
different Higgs search channels are essentially free from these uncertainties
and do not require further theoretical assumptions, on the total Higgs decay
width for instance.  These ratios, in particular in the $H\to \gamma \gamma$
v.s. $H\to ZZ$  and $H\to \tau \tau$ v.s. $H\to WW$ channels, are being
measured quite accurately already with the present data and provide tests  of
the SM predictions in a less model--dependent way. We show that at the 14 TeV
LHC with a high luminosity, 300 fb$^{-1}$ and even 3000 fb$^{-1}$, they
could allow the measurement of ratios of Higgs couplings with an accuracy at
the level of a  few percent which should allow to test the small deviations
expected  in realistic  new physics models. 

In a second part of this paper, we have considered together with the ratios
of  cross sections times branching ratios in the most important search
channels,  the signal strength in the extremely clean $H\to ZZ$ channel in which
the theoretical  uncertainty is taken  to be a bias. We have then shown that
first, the particle observed at the LHC is at most 68\% CP--odd at the
99\%CL and  the possibility that it is a pure pseudoscalar state (and hence does not
couple to $ZZ$ states at tree--level) is excluded at the $4\sigma$ level when
including both the  experimental and theoretical uncertainties. The signal
strengths in the $H\to ZZ$ channel also measure the invisible Higgs decay
width which is shown to be  $\Gamma_H^{\rm inv}/  \Gamma_H^{\rm SM} \leq 0.52$ 
at the 68\%CL if the Higgs couplings to fermions and gauge bosons are
assumed to be SM--like.

All these results give us great confidence that the state observed at the LHC in
July 2012 is indeed a Higgs particle and, more than that, it resembles  very
closely to the Higgs particle predicted in the Standard Model.\bigskip   

\noindent {\bf Acknowledgements}:  We thank Aleksandr Azatov, Henri Bachacou, Oscar J.~P.~Eboli and
Rohini Godbole for discussions. AD thanks the CERN theory division for the kind
hospitality and GM the University of Warsaw where this work was finalized.
This work is  supported by the ERC Advanced Grant Higgs@LHC. 
GM is also supported by the ``Institut Universitaire de France'',  the
ANR {CPV-LFV-LHC}  project and the Orsay Labex P2IO.

\end{document}